\documentclass{atmp-7x10}
\usepackage{graphicx}
\usepackage{cite}
\usepackage{amsmath}
\usepackage{amsfonts}
\usepackage{amssymb}
\makeatletter
\def\footnote{\stepcounter{totfootns}\@ifnextchar[\@xfootnote{\stepcounter\@mpfn
     \protected@xdef\@thefnmark{\thempfn}%
     \h@footnotemark\!\!\h@footnotetext}}
\makeatother

\graphicspath{{H://IP/ATMP/ISSUE/13-1/Pagination/art/}}

\newcommand{\lC}{\mathrm{l\hspace{-2.1mm}C}}
\newcommand{\lR}{\mathrm{I\hspace{-0.7mm}R}}

\numberwithin{equation}{section}

\begin{document}

\title[K\"ahler--Ricci flow, Morse theory]{K\"ahler--Ricci flow, Morse theory,
and vacuum structure deformation of $N=1$
supersymmetry\break in four dimensions}

\author[Bobby E. Gunara and Freddy P. Zen]{Bobby E. Gunara and Freddy P.
Zen}

\copyrightnotice{2009}{13}{217}{257}

\arxurl{hep-th/0708.1036}

\address{Indonesia Center for Theoretical and Mathematical
Physics (ICTMP) and~Theoretical  Physics Laboratory,
Theoretical High Energy Physics and Instrumentation
Division, Faculty of Mathematics and Natural Sciences,
Institut Teknologi Bandung, Jl. Ganesha 10, Bandung 40132,
Indonesia}
\addressemail{bobby@fi.itb.ac.id, fpzen@fi.itb.ac.id}

\begin{abstract}
We address some aspects of four-dimensional chiral $N=1$
supersymmetric theories on which the scalar manifold is
described by K\"ahler geometry and can further be viewed as
K\"ahler--Ricci soliton generating a one-parameter family
of K\"ahler geometries. All couplings
 and solutions, namely  the BPS domain walls and their supersymmetric
 Lorentz invariant vacua turn out to be evolved with respect
 to the flow parameter related to the soliton. Two models are discussed, namely $N=1$ theory
 on K\"ahler--Einstein manifold and $U(n)$ symmetric
 K\"ahler--Ricci
 soliton with positive definite metric. In the first case, we find that the
evolution of the soliton causes topological change and
correspondingly,  modifies the\hfill{\break}
\end{abstract}

\maketitle

\begin{center}
\begin{quote}\leftskip=0.6pc\rightskip=-0.1pc
{\fontsize{10}{10.95}\selectfont  Morse index of the
nondegenerate vacua realized in the parity transformation
of the Hessian matrix of the scalar potential after hitting
singularity, which is natural in the global theory and for
nondegenerate Mink\-owskian vacua of the local theory.
However, such situation is not trivial in anti de Sitter
vacua. In an explicit model, we find that this geometric
(K\"ahler--Ricci) flow can also change the index of the
vacuum before and after singularity. Finally in the second
case, since around the origin the metric is diffeomorphic
to $\;{\mathrm{\lC P}}^{n -1}$, we have to consider it in
the asymptotic region. Our analysis shows that no index
modification of vacua is present in both global and local
theories.}
\end{quote}
\end{center}

\section{Introduction}

Geometric evolution equation which is called Ricci flow
equation introduced by Hamilton in \cite{Ham} has been
prominently studied by mathematicians due to its
achievement in solving famous three-dimensional puzzles,
namely Poincar\'e and Thurston's geometrization conjectures
\cite{Per}.\footnote{For a detailed and comprehensive proof
of these conjectures is given in \cite{caozhu,KL,MT}.} \nobreak
Moreover,~in~higher dimensional manifold, particularly in
compact K\"ahler manifold with the first Chern class $c_1
=0$ or $c_1 <0$, one can then reprove the well-known Calabi
conjecture using the so called K\"ahler--Ricci flow
equation \cite{cao0}. Several authors have studied to
construct a soliton of the K\"ahler--Ricci flow with $U(n)$
symmetry \cite{cao,cao1,FIK}. Such a solution is one of our
interest and related to the subject of this paper.

In the physical context, this K\"ahler--Ricci flow can be
regarded as a one-loop approximation to the physical
renormalization group equation of two-dimensional theories
at quantum level. For example, some authors have discussed
it in the context of Wilsonian renormalization group coming
from $N=2$ supersymmetry in two and three dimensions
\cite{HI, MN}.\footnote{We thank M. Nitta for informing us
these papers.} However this is not the case in higher
dimension, particularly in four dimensions.

The purpose of this paper is to study the nature of chiral
$N=1$ supersymmetric theories in four dimensions together
with its solutions, when the scalar manifold described by
the K\"ahler manifold is no longer static.  In the sense
that it is a K\"ahler--Ricci soliton generating a
one-parameter family of K\"ahler manifolds which evolves
with respect to a parameter, say $\tau$. This further imply
that all couplings such as scalar potential, fermion mass
matrix, and gravitino mass do have evolution with respect
to the flow parameter~$\tau$.

Particularly in local theory such deformation
 happens in its solitonic solution, like BPS domain walls that
have residual supersymmetry. This can be easily seen
because all quantities such as warped factor, BPS
equations, and the beta function of field theory in the
context of anti de Sitter (AdS)/conformal field theory
(CFT) correspondence (for a review, see for
example~\cite{AdSCFT}) do depend on $\tau$. In this paper,
we carry out the analysis on critical points of the scalar
potential describing Lorentz invariant vacua on which the
BPS equations and the beta function vanish. The first step
is to consider the second-order analysis using Hessian
matrix of the scalar potential which gives the index of the
vacua. The next step is to perform a first-order analysis
of the beta function for verifying the existence of an AdS
vacuum which corresponds to a CFT in three dimensions. We
realize our analysis into two models as follows.

In the first model we have a global and local $N=1$
supersymmetric theories on the soliton where the initial
manifold is K\"ahler--Einstein whose ``cosmological
constant'' is chosen, for example, to be positive, namely
$\Lambda
>0$. By taking $\tau \ge 0$ the flow turns out to be
collapsed to a point at $\tau = 1/2\Lambda$ and
correspondingly, both global and local $N=1$ theories blow
up. Furthermore, this geometric flow interpolates between
two different theories which are disconnected by the
singularity at $\tau = 1/2\Lambda$, namely $N=1$ theories
on K\"ahler--Einstein manifold with $\Lambda >0$ in the
interval $0 \le \tau < 1/2\Lambda$, while for $\tau >
1/2\Lambda$ it becomes $N=1$ theories on K\"ahler--Einstein
manifold with $\Lambda < 0$ and opposite metric signature
which yields a theory with wrong-sign kinetic terms and
therefore, might be considered as ``ghosts'' at the quantum
level.\footnote{This aspect will be addressed elsewhere.}

Consequently, in global $N=1$ chiral theory such evolution
of\break \hbox{K\"ahler--Einstein} soliton results in the
deformation of the supersymmetric Minkow\-skian vacua which
can be immediately observed from the parity transformation
of the Hessian matrix of the scalar potential that changes
the index of the vacua if they are nondegenerate. In local
theory, we get the same situation for nondegenerate
Minkowskian vacua. These happen because only the geometric
(K\"ahler--Ricci) flow affects the index of the vacua in
the second-order \hbox{analysis.}

In AdS ground states we have however a different situation.
In this case both
 the geometric flow and the coupling quantities  such as the holomorphic
  superpotential  and the fermionic mass do play a role in determining
  the index of the vacua. Therefore, we have to enforce certain
  conditions on the Hessian matrix of the scalar potential
  in order to obtain the parity transformation. In addition,
 there is a possibility of having a condition where the degeneracy of
 the vacua can also be modified by the K\"ahler--Ricci flow. For example,
 in $\;{\mathrm{\lC P}}^1$ model with linear superpotential, we
 find that this geometric flow transforms a nondegenerate vacuum
 to a degenerate vacuum and vice versa with respect to $\tau$ before
 and after singular point at $\tau = 1/4$. In this model,
 degenerate vacuum possibly could not exist particularly in the
 infrared (IR) (low energy scale) region.

Now we turn to consider the second model, namely $N=1$
theories on $U(n)$ invariant K\"ahler--Ricci soliton
constructed over line bundle of $\;{\mathrm{\lC P}}^{n -1}$
with positive definite metric for $\tau \ge 0$ \cite{FIK}.
Near the origin, the soliton turns to the
K\"ahler--Einstein manifold, namely $\;{\mathrm{\lC P}}^{n
-1}$. Hence, we have to analyze its behavior in the
asymptotic region. By taking $\Lambda >0$ case, we find
that the soliton collapses into a K\"ahler cone at $\tau =
1/2\Lambda$ which is singular. Moreover, the soliton is
asymptotically dominated by the cone metric which is
positive definite for all $\tau  \ge 0$. As a result the
vacuum is static, or in other words it does not depend on
$\tau$. Therefore, no parity transformation of the Hessian
matrix and no other vacuum modification caused by the
geometric flow~exist.

\enlargethispage{6pt} The organization of this paper is as
follows. In Section \ref{EEKRF} we introduce the notion of
K\"ahler--Ricci flow equation with its solutions. Section
\ref{STKS} consists of two part discussions about the
global and local $N=1$ supersymmetry on general
K\"ahler--Ricci soliton. Then, we construct BPS domain wall
solutions in Section \ref{BPSsec}. In Section \ref{SUSYKEM}
we discuss the behavior of the supersymmetric Lorentz
invariant vacua on K\"ahler--Einstein manifold and gives an
explicit model. Section \ref{SUSYKRS} provides a review of
$U(n)$ symmetric K\"ahler--Ricci soliton together with our
analysis on its vacua. We finally conclude our results in
Section~\ref{concl}.

\section{Evolution equation: K\"ahler--Ricci flow}\label{EEKRF}

This section is devoted to give a background about Ricci
flow defined on a complex manifold endowed with K\"ahler
metric which is called K\"ahler--Ricci flow. A remarkable
fact is that the solution of this K\"ahler--Ricci flow
remains K\"ahler as demanded by $N=1$ supersymmetry. There
are some excellent references on this subject for
interested reader, for example, in
\cite{cao,cao1,FIK}.\footnote{We also recommend
\cite{topping} for a detailed look on Ricci flow.} We
collect some evolution equations of Riemann curvature,
Ricci tensor, and Ricci scalar in Appendix~C.

A K\"ahler manifold $({\bf{M}},g(\tau))$ is a
K\"ahler--Ricci soliton if it satisfies
\begin{equation}\label{KRF}
\frac{\partial g_{i\bar{j}}}{\partial \tau}(z, \bar{z};
\tau) = -2 R_{i\bar{j}}(z, \bar{z}; \tau),\quad 0 \le \tau
<T,
\end{equation}
 where $(z, \bar{z}) \in {\bf{M}}$. Furthermore, the flow
 $g(\tau)$ has the form
\begin{equation}\label{KRsolumum}
g(\tau) = \sigma(\tau) \psi^*_{\tau}(g(0)), \quad 0 \le
\tau < T,
\end{equation}
with $\sigma(\tau) \equiv (1-2\Lambda \tau)$, the map
$\psi_{\tau}$ is a diffeomorphism, and  $g(0)$ is the
initial metric at $\tau =0$ which fulfills the following
identity
\begin{equation}\label{initialgeom}
-2R_{i\bar{j}}(0) = \nabla_i Y_{\bar{j}}(0)+
\bar{\nabla}_{\bar{j}} Y_i(0) -2\Lambda g_{i\bar{j}}(0),
\end{equation}
for some real constant $\Lambda$ and some holomorphic
vector fields\footnote{Holomorphicity of $Y(0)$ follows
from the fact that the complex structure $J$ on ${\bf{M}}$
satisfies $\psi^*_{\tau}(J)= J$.}
\begin{equation}\label{killdiff0}
Y(0) = Y^i(z, 0) \partial_i + Y^{\bar{i}}(\bar{z}, 0)
\bar{\partial}_{\bar{i}},
\end{equation}
 on ${\bf{M}}$ where $i,j = 1,\ldots, {\text{dim}}_{\:\mathrm{l\hspace{-1.6mm}C}}({\bf{M}})$.
 The vector field
$Y(0)$ is related to a $\tau$-dependent vector field $X(\tau)$ by
\begin{equation}\label{killdiff}
X(\tau) = \frac{1}{\sigma(\tau)}Y( 0),
\end{equation}
which generates a family of diffeomorphisms $\psi_{\tau}$
and satisfies the following equation
\begin{align}
\frac{\partial \hat{z}^i}{\partial \tau} &= X^i(\hat{z}, \tau),\nonumber\\
\frac{\partial \hat{\bar{z}}^{\bar{i}}}{\partial \tau} &=
X^{\bar{i}}(\hat{\bar{z}}, \tau),\label{diffeo}
\end{align}
with
\begin{equation}\label{diffeo1}
\hat{z} \equiv \psi_{\tau}(z).
\end{equation}

\enlargethispage{6pt} \subsection{K\"ahler--Einstein
manifold}

We turn to consider the case where $Y(0)$ vanishes. In this
case, the initial metric is K\"ahler--Einstein
\begin{equation}\label{KEmet}
R_{i\bar{j}}(0) = \Lambda \,g_{i\bar{j}}(0),
\end{equation}
for some constants $\Lambda \in \lR$. Taking
\begin{equation}
g_{i\bar{j}}(\tau) = \rho(\tau) \,g_{i\bar{j}}(0),
\end{equation}
and then from the definition of Ricci tensor, one finds that
\begin{equation}\label{RiccKEM}
R_{i\bar{j}}(\tau) = R_{i\bar{j}}(0) = \Lambda
\,g_{i\bar{j}}(0).
\end{equation}
So the solution of (\ref{KRF}) can be obtained as
\begin{equation}\label{solKEM}
g_{i\bar{j}}(\tau) = (1-2\Lambda \tau) \,g_{i\bar{j}}(0),
\end{equation}
with $\rho(\tau) = \sigma(\tau) = (1-2\Lambda \tau)$.
Moreover, the K\"ahler potential has the~form
\begin{equation}\label{KpotKEM}
K(\tau) = (1-2\Lambda \tau) \, K(0).
\end{equation}

If we take the K\"ahler--Ricci flow (\ref{KRF}) to evolve
 for $\tau \ge 0$, then for the case $\Lambda < 0$ the
metric (\ref{solKEM}) is smoothly expanded and no collapsing flow
exists. On the other hand, in the case $\Lambda > 0$ it collapses
to a point at $\tau = 1/2\Lambda$, but reappear for $\tau >
1/2\Lambda$ which is diffeomorphic to a manifold with $\Lambda <
0$ and
 its metric signature is opposite with the initial metric defined
in (\ref{KEmet}).\footnote{See discussion in Sections
\ref{GSC} and \ref{LSC}.} Note that in both cases the flow
also becomes singular as $\tau \to +\infty$. In addition,
if the initial metric is Ricci flat, namely $\Lambda =0$,
then it is clear that the metric does not change
under~(\ref{KRF}).

\subsection{Gradient K\"ahler--Ricci soliton}

Now let us discuss the $Y(0) \ne 0$  case. Our particular
interest is the case where the holomorphic vector field
(\ref{killdiff}) can also be written~as
\begin{equation}\label{gradvec}
Y^i(0) = g^{i\bar{j}}\bar{\partial}_{\bar{j}}P(z, \bar{z})
\end{equation}
for some real functions $P(z, \bar{z})$ on ${\bf{M}}$.
Thus,  we say that $g(\tau)$ is a gradient K\"ahler--Ricci
soliton \cite{cao, cao1}. In this case,  in general, there
are three possible solutions. For $\Lambda
>0$ we have a shrinking gradient K\"ahler--Ricci soliton, whereas
 the soliton is an expanding gradient K\"ahler--Ricci soliton for $\Lambda <0$.
 The case where $\Lambda =0$ is a steady gradient K\"ahler--Ricci soliton.
 In fact, on any compact K\"ahler manifold, a gradient steady or expanding
 K\"ahler--Ricci soliton is necessarily a K\"ahler--Einstein manifold because
 the real function $P(z, \bar{z})$ is just a constant \cite{caozhu}.
 We give an explicit construction of these expanding and shrinking
 K\"ahler--Ricci solitons in Section~\ref{SUSYKRS}.

\section{$N=1$ supersymmetric theory on K\"ahler--Ricci soliton}\label{STKS}

In this section we discuss $N=1$ supersymmetric theory in
four dimensions whose nonlinear $\sigma$-model describing a
K\"ahler geometry\footnote{Note that in four dimensional
global $N=1$ supersymmetry the scalar manifold $\bf{M}$ is
K\"ahler, whereas local $N=1$ supersymmetry demands that
the scalar manifold $\bf{M}$ has to be Hodge--K\"ahler
\cite{wess, DF}, see the next subsection.}
$({\bf{M}},g(\tau))$ satisfies one-parameter
K\"ahler--Ricci flow equation (\ref{KRF}). This equation
tells us that the dynamics of the metric $g_{i\bar{j}}$
with respect to $\tau$ is determined by the Ricci tensor
$R_{i\bar{j}}$. Thus, it follows that the $N=1$ theory is
deformed with respect to $\tau$ as discussed in the
following. We divide the discussion into two parts. In the
first part we construct a global $N=1$ chiral supersymmetry
as a toy model. Then, in the second part we extend the
construction to a local supersymmetry, namely $N=1$ chiral
supergravity which corresponds to our analysis for the most
part of this~paper.

\subsection{Global $N=1$ chiral supersymmetry}

First of all, let us focus on the properties of the
deformed global $N=1$ chiral supersymmetric theory in four
dimensions. The spectrum of the theory consists of a
spin-${\frac{1}{2}}$ fermion $\chi^i$ and a complex scalar
$z^i$ with $i,j = 1,\ldots,n_c$. As mentioned above, the
complex scalars $(\bar{z}^{\bar{i}},z^i)$ parameterize a
K\"ahler  geometry $({\bf{M}},g(\tau))$. The construction
of the global $N=1$ theory on K\"ahler--Ricci soliton is as
follows. First, we consider the chiral Lagrangian in (for a
pedagogical review of $N=1$ supersymmetry, see for
example~\cite{wess}), say
${\mathcal{L}}^0_{\mathrm{global}}$, where the metric of
the scalar manifold is static. Then, replacing all
geometric quantities such as the metric $g_{i\bar{j}}(0)$
by the soliton $g_{i\bar{j}}(\tau)$, the on-shell $N=1$
chiral Lagrangian can be written down up to 4-fermion
term~as
\begin{align}
{\mathcal{L}}_{\mathrm{global}} &=
g_{i\bar{j}}(\tau)\,\partial_{\mu} z^i
\partial^{\mu}\bar{z}^{\bar{j}} -\frac{\mathrm{i}}{2}g_{i\bar{j}}(\tau)
 \left(\bar{\chi}^{i} \gamma^\mu
\partial_\mu \chi^{\bar{j}} + \bar{\chi}^{\bar{j}} \gamma^\mu
  \partial_\mu \chi^{i}\right) \nonumber\\
  &\quad+ M_{ij}(\tau)\bar{\chi}^i \chi^j
  + \bar{M}_{\bar{i}\bar{j}}(\tau)\ \bar{\chi}^{\bar{i}} \chi^{\bar{j}}
  - V(\tau),\label{globL}
\end{align}
where the scalar potential has the form
\begin{equation}\label{V}
 V(z,\bar{z};\tau) =   g^{i\bar{j}}(\tau)\, \partial_i W
\,\bar{\partial}_{\bar{j}}\bar{W},
\end{equation}
with $W \equiv W(z)$. The metric $g_{i\bar{j}}(\tau) \equiv
\partial_i \bar{\partial}_{\bar{j}}K(\tau)$ is the solution of
(\ref{KRF}), where the real function $K(\tau)$ is a
K\"ahler potential. Fermionic mass-like quantity
$M_{ij}(\tau)$ is then given~by
\begin{equation}\label{fermass}
M_{ij}(\tau) \equiv \nabla_i \partial_j W = \partial_i
\partial_j W - \Gamma^k_{ij}(\tau)\, \partial_k W.
\end{equation}
and it is related to the first derivative of the scalar potential
(\ref{V}) with respect to $(z, \bar{z})$
\begin{align}
\partial_i V &=  g^{j\bar{l}}(\tau)
M_{ij}(\tau) \, \bar{\partial}_{\bar{l}}\bar{W},
\nonumber\\
\partial_{\bar{i}} V &=
 g^{\bar{j}l}(\tau)
\bar{M}_{\bar{i}\bar{j}}(\tau)
\partial_l W. \label{dVz}
\end{align}
The supersymmetry transformation of the fields up to
3-fermion terms leaving the Lagrangian invariant
(\ref{globL})~are
\begin{align}
\delta z^i &= \bar{\chi}^{i} \epsilon_1,\nonumber\\
\delta \chi^{i} &= {\mathrm{i}}
\partial_{\mu}z^i\gamma^{\mu}\epsilon^1 +
g^{i\bar{j}}(\tau)
 \bar{\partial}_{\bar{j}}\bar{W}\epsilon_1 \label{susyvar}.
\end{align}
Since the evolution of the metric $g_{i\bar{j}}(\tau)$ is
determined by the Ricci tensor $R_{i\bar{j}}(\tau)$, one
can directly see that the dynamics of the scalar potential
(\ref{V}) and the fermionic mass (\ref{fermass}) has a
similar behavior with respect to $\tau$, namely
\begin{align}
 \frac{\partial V(\tau)}{\partial \tau} &= 2 R^{i\bar{j}}(\tau)\,
  \partial_i W \,
 \bar{\partial}_{\bar{j}}\bar{W}, \nonumber\\
\frac{\partial M_{ij}(\tau)}{\partial \tau} &=  2
g^{k\bar{l}}(\tau) \nabla_i R_{j\bar{l}}(\tau)\, \partial_k
W, \label{dVt}
\end{align}
where $R^{i\bar{j}} \equiv g^{i\bar{l}} g^{k\bar{j}}R_{k\bar{l}}$.

To make the above formula clearer, we now turn to discuss
an example where the initial geometry (at $\tau =0$) is
K\"ahler--Einstein manifold, as discussed in section
\ref{EEKRF}. This follows that the equations in (\ref{dVt})
are simplified~into
\begin{align}
 \frac{\partial V(\tau)}{\partial \tau} &= 2 \Lambda \, \sigma(\tau)^{-2}\,
 g^{i\bar{j}}(0) \, \partial_i W \,
 \bar{\partial}_{\bar{j}}\bar{W} = 2 \Lambda \, \sigma(\tau)^{-2}\,
 V(0), \nonumber\\
\frac{\partial M_{ij}(\tau)}{\partial \tau} &=
0.\label{dVtKE}
\end{align}
The first equation in (\ref{dVtKE}) has the solution
\begin{equation}\label{VKE}
V(\tau) = \sigma(\tau)^{-1} \, V(0),
\end{equation}
whereas the second equation tells us that the fermion mass
matrix $M_{ij}(\tau)$ does not depend on $\tau$. As the
flow (\ref{solKEM}) deforms for $\tau \ge 0$, the
Lagrangian (\ref{globL}) can be smoothly defined and there
is no singularity for the case $\Lambda < 0$. This follows
that in this case we have a well-defined theory as the
K\"ahler geometry changed with respect to $\tau$.  However,
in the case $\Lambda > 0$ since the flow (\ref{solKEM})
shrinks to a point at $\tau = 1/2\Lambda$ the scalar
potential (\ref{VKE}) becomes singular and the Lagrangian
(\ref{globL}) diverges. Moreover, as mentioned above for
$\tau > 1/2\Lambda$, the flow evolves again and it is
related to another $N=1$ global theory on a
K\"ahler--Einstein manifold with $\Lambda < 0$ and
different metric signature.

To be precise, let ${\bf M}_0$ be the initial geometry
which is K\"ahler--Einstein manifold with $\Lambda
> 0$ whose metric $g_{i\bar{j}}(0)$ is positive definite.
Next, we cast (\ref{RiccKEM}) to the following~form
\begin{equation}\label{RiccKEM1}
R_{i\bar{j}}(\tau) = \widehat{\Lambda}(\tau) \,
g_{i\bar{j}}(\tau),
\end{equation}
where
\begin{align}
 \widehat{\Lambda}(\tau) &\equiv \sigma(\tau)^{-1} \Lambda,
 \nonumber\\
 g_{i\bar{j}}(\tau) &\equiv\sigma(\tau) \,
 g_{i\bar{j}}(0).
\end{align}
In the interval $0 \le \tau < 1/2\Lambda$, the solution is
diffeomorphic to the initial geometry. However, for $\tau >
1/2\Lambda$ we have a K\"ahler--Einstein geometry ${\bf
\widehat{M}}_0$ with a negative definite metric
$g_{i\bar{j}}(\tau)$ and $\widehat{\Lambda}(\tau) < 0$. As
mentioned in the introduction, such a metric produces a
theory
with wrong-sign kinetic terms and could be interpreted as ghosts.

Similar result can also be achieved for the case where the
initial geometry is a K\"ahler--Einstein geometry with
indefinite metric where $\Lambda > 0$ \cite{Petean}. Then,
we have also an indefinite K\"ahler--Einstein metric but
with $\widehat{\Lambda} < 0$ for $ \tau > 1/2\Lambda$.

\subsection{$N=1$ chiral supergravity}

In this subsection, we generalize the previous result to a
local $N=1$ supersymmetry. In four dimensions the spectrum
of a generic chiral $N=1$ supergravity theory consists of a
gravitational multiplet and $n_c$ chiral multiplets. These
multiplets are decomposed of the following component
fields:
\begin{itemize}
\item A gravitational multiplet
\begin{equation}
 (g_{\mu\nu}, \psi^1_{\mu}), \quad \mu =0,\ldots,3.
\end{equation}
This multiplet consist of the graviton $g_{\mu\nu}$ and a
gravitino $\psi^1_{\mu}$. For the gravitino
$\psi^1_{\mu},\psi_{1\mu}$ and the upper or lower index
denotes left or right chirality, respectively.

\item $n_c$ chiral multiplets
\begin{equation}
(z^i, \chi^i ), \quad i = 1,\ldots,n_c.
\end{equation}
Each chiral multiplet consist of a spin-${\frac{1}{2}}$ fermion
$\chi^i$ and a complex scalar $z^i$.
\end{itemize}
Local supersymmetry further requires the scalar manifold
$\bf{M}$ spanned by the complex scalars
$(\bar{z}^{\bar{i}},z^i)$ to be a Hodge--K\"ahler manifold
with an additional $U(1)$-connection
\begin{equation}\label{U1con}
Q \equiv -\frac{1}{M_P^2}\left( K_i \,dz^i- K_{\bar{i}}\,
d\bar{z}^{\bar{i}}\right)\!,
\end{equation}
where $K_i \equiv \partial_i K$ and the metric $g_{i\bar{j}} =
\partial_i \bar{\partial}_{\bar{j}}K$
 where the K\"ahler potential $K$ is an arbitrary real function
  \cite{wess,DF}.
 In our case since $g_{i\bar{j}}$ is the solution of (\ref{KRF}),
  then $K$ and $Q$ must be $\tau$-dependent.

Similar as in the global case, we first consider the chiral
theory studied in \cite{wess,DF}. Then, by employing the
same procedure the $N=1$
\hbox{supergravity}\vadjust{\pagebreak} Lagrangian up to
four-fermions terms on K\"ahler--Ricci soliton has
the\break following form
\begin{align}
 {\mathcal{L}}_{\mathrm{local}} &= -\frac{M_P^2}{2}R +
g_{i\bar{j}}(\tau)\partial_{\mu}
z^i\partial^{\mu}\bar{z}^{\bar{i}} +
\frac{\epsilon^{\mu\nu\lambda\sigma}}{\sqrt{-h}}\left(\bar{\psi}^1_\mu
  \gamma_{\sigma}\widetilde{\nabla}_{\nu}\psi_{1\lambda}-\bar{\psi}_{1\mu}
  \gamma_{\sigma}\widetilde{\nabla}_{\nu}\psi^1_{\lambda}\right)\nonumber\\
&\quad- \frac{\mathrm{i}}{2}
g_{i\bar{j}}(\tau)\left(\bar{\chi}^{i} \gamma^\mu
\nabla_\mu \chi^{\bar{j}} + \bar{\chi}^{\bar{j}} \gamma^\mu
  \nabla_\mu \chi^{i}\right) \nonumber\\
  &\quad- g_{i\bar{j}}(\tau)
  \left(\bar{\psi}_{1\nu}\gamma^{\mu}\gamma^{\nu}\chi^i
  \partial_{\mu}\bar{z}^{\bar{j}} +
  \bar{\psi}_{\nu}^1\gamma^{\mu}\gamma^{\nu}\chi^{\bar{j}}\partial_{\mu}z^i
  \right)\nonumber\\
&\quad+ L(\tau)\,
\bar{\psi}^1_{\mu}\gamma^{\mu\nu}\psi^1_{\nu} +
\bar{L}(\tau)\, \bar{\psi}_{1\mu}\gamma^{\mu\nu}\psi_{1\nu}
+ {\mathrm{i}} g_{i\bar{j}}(\tau)\notag\\
&\quad\times\left(\bar{N}^{\bar{j}}(\tau)\bar{\chi}^i
\gamma^\mu \psi_{\mu}^1 + N^i(\tau)
\bar{\chi}^{\bar{j}}\gamma^\mu
\psi_{1\mu}\right)\nonumber\\
&\quad+ {\mathcal{M}}_{ij}(\tau)\bar{\chi}^i \chi^j +
\bar{\mathcal{M}}_{\bar{i}\bar{j}}(\tau)\
\bar{\chi}^{\bar{i}} \chi^{\bar{j}} -
{\mathcal{V}}(\tau),\label{L1}
\end{align}
where $h \equiv {\mathrm{det}}(g_{\mu\nu})$ and
$L(\tau)(\bar{L}(\tau))$ can be written in terms of an
(anti)-holomorphic superpotential function of
$W(z)(\bar{W}(\bar{z}))$,
\begin{align}
L(\tau) &= e^{K(\tau)/2M_P^2}W(z),\nonumber\\
\bar{L}(\tau) &=
e^{K(\tau)/2M_P^2}\bar{W}(\bar{z}),\label{gravitinomass}
\end{align}
with  $K(\tau)$ is a K\"ahler potential of the chiral
multiplets, and the quantities
$N^i(\tau),{\mathcal{M}}_{ij}(\tau)$ are given~by:
\begin{align}
N^i(\tau) &=  g^{i\bar{j}}(\tau)
\bar{\nabla}_{\bar{j}}\bar{L}(\tau),\nonumber\\
{\mathcal{M}}_{ij}(\tau) &= \frac{1}{2}\nabla_i \nabla_j
L(\tau). \label{fermass1}
\end{align}
On the other hand, the $N=1$ scalar potential can be expressed in
terms of $L(\bar{L})$ as
\begin{equation}\label{V1}
{\mathcal{V}}(\tau) = g^{i\bar{j}}(\tau) \nabla_i
L(\tau)\,\bar{\nabla}_{\bar{j}}\bar{L}(\tau) -
\frac{3}{M_P^2} L(\tau)\bar{L}(\tau),
\end{equation}
where $\nabla_i L(\tau) = \partial_i L(\tau) +
\frac{1}{2M_P^2}K_i(\tau) \,L(\tau)$. Then the first derivative of
${\mathcal{V}}(\tau)$ with respect to $(\bar{z}^{\bar{i}},z^i)$
are of the form
\begin{align}
\partial_{\bar{i}}{\mathcal{V}} &= \frac{1}{2}
{\mathcal{M}}_{kj}(\tau) N^j(\tau)- \frac{1}{M_P^2}
N_k(\tau)\bar{L}(\tau),
\nonumber\\
\partial _{\bar{i}} {\mathcal{V}}  &=
\frac{1}{2}\bar{{\mathcal{M}}}_{\bar{k}\bar{j}}(\tau)\bar{N}^{\bar{j}}
(\tau)-\frac{1}{M_P^2}\bar{N}_{\bar{k}}(\tau) L(\tau).
\end{align}
The supersymmetry transformation laws up to 3-fermion terms
leaving\break \hbox{invariant} (\ref{L1})~are:
\begin{align}
\delta \psi_{1 \mu} &= M_P \left(
\widetilde{{\mathcal{D}}}_\mu \epsilon_1 +
\frac{\mathrm{i}}{2}L(\tau)\,
\gamma_\mu\, \epsilon^1 \right)\!,\nonumber\\
\delta \chi^{i} &= {\mathrm{i}}
\partial_{\mu}z^i\gamma^{\mu}\epsilon^1 + N^i(\tau)
\epsilon_1, \label{susyvar1}\\
\delta e^a_\mu &= -  \frac{1}{M_P} (\mathrm{i}\bar{\psi}_{1
\mu}\gamma^a \epsilon^1 +
\mathrm{i}\bar{\psi}^{1 \mu}\gamma^a \epsilon_1 ),\nonumber\\
\delta z^i &= \bar{\chi}^{i} \epsilon_1,\nonumber
\end{align}
where $\widetilde{{\mathcal{D}}}_\mu \epsilon_1 =
\partial_{\mu}\epsilon_1
-\frac{1}{4}\gamma_{ab}\,\omega^{ab}_{\mu}\epsilon_1+\frac{\mathrm{i}}{2}
Q_{\mu}(\tau)\epsilon_1$, and $Q_{\mu}(\tau)$ is a
$U(1)$-connection which is the first derivative of the
K\"ahler potential $K(\tau)$ with respect to $x^{\mu}$.
Here, we have defined $\epsilon_1 \equiv \epsilon_1(x,
\tau)$. Furthermore the evolution of the coupling
quantities beside the metric $g_{i\bar{j}}(\tau)$ in
(\ref{L1}) are given~by
\begin{align}
\frac{\partial L(\tau)}{\partial \tau} &=
\frac{K_{\tau}(\tau)}{2M_P^2}
 L(\tau),\nonumber\\
\frac{\partial N^i(\tau)}{\partial \tau} &=  2 R^i_j(\tau)
N^j(\tau)+ \frac{K_{\tau}(\tau)}{2M_P^2} N^i(\tau)+
g^{i\bar{j}}(\tau) \frac{K_{\bar{j}\tau}(\tau)}{M_P^2}
\bar{L}(\tau), \nonumber\\
 \frac{\partial {\mathcal{V}}(\tau)}{\partial
\tau} &= \frac{\partial N^i(\tau)}{\partial \tau}
N_i(\tau)+ \frac{\partial N_i(\tau)}{\partial \tau}
N^i(\tau) -
\frac{3K_{\tau}(\tau)}{M_P^2} \vert L(\tau)\vert^2, \label{dV1t}\\
\frac{\partial {{\mathcal{M}}_{ij}(\tau)}}{\partial \tau}
&= -R_{j\bar{l}}(\tau) \nabla_i \bar{N}^{\bar{l}}(\tau)\notag\\
&\quad+ \frac{1}{2} g_{j\bar{l}}(\tau) \left( \partial_i
\frac{\partial \bar{N}^{\bar{l}}(\tau)}{\partial \tau}+
\frac{K_{i\tau}(\tau)}{2M_P^2}
 \bar{N}^{\bar{l}}(\tau)+  \frac{K_i(\tau) }{2M_P^2}
\frac{\partial \bar{N}^{\bar{l}}(\tau)}{\partial \tau}
\right)\!.\nonumber
\end{align}

Similar as in the global case, we finally take an example
where the initial metric is K\"ahler--Einstein. In this
case the metric evolves with respect to $\tau$ as in
(\ref{solKEM}) and the $U(1)$-connection takes the~form
\begin{equation}\label{U1conKE}
Q(\tau) \equiv - \frac{\sigma(\tau)}{M_P^2}\left(
\partial_i K(0) dz^i- \bar{\partial}_{\bar{i}}K(0)
d\bar{z}^{\bar{i}}\right)\!,
\end{equation}
since the K\"ahler potential is given by (\ref{KpotKEM}).
Taking $\tau \ge 0$, the theory is well defined for
$\Lambda < 0$, while for $\Lambda > 0$ it becomes singular
at $\tau = 1/2\Lambda$. In the latter case, namely $\Lambda
> 0$ case, the $N=1$ local theory is changed and the shape
of the scalar manifold is a K\"ahler--Einstein manifold
with $\Lambda < 0$ and moreover, the metric signature has
opposite sign. Thus, we have a gravitational multiplet
coupled to chiral multiplets which might be considered as
ghosts.

\section{Deformation of $N=1$ supergravity BPS domain walls}\label{BPSsec}

Here, we turn our attention to discuss a ground state which
breaks Lorentz invariance partially  and preserves half of
the supersymmetry of the parental theory, namely the BPS
domain wall in $N=1$ supergravity. This type of solution
was firstly discovered in \cite{BPSDW}. Interested reader
can further consult \cite{RevBPSDW} for an excellent review
of this subject. In this section some conventions follow
rather closely to \cite{CDGKL,GZA}.

The first step is to take the ansatz metric as
\begin{equation}\label{DWans}
ds^2 = a^2(u,
\tau)\,\eta_{\underline{\nu}\,\underline{\lambda}}\,dx^{\underline{\nu}}\,dx^{\underline{\lambda}}-du^2,
\end{equation}
where $\underline{\nu}, \underline{\lambda}=0,1,2$,
$\eta_{\underline{\nu}\,\underline{\lambda}}$ is a
three-dimensional Minkowskian metric, and then the
corresponding Ricci scalar has the~form
\begin{equation}\label{Rans}
 R = 6 \left[ \left( \frac{a'}{a} \right)'
 + 2 \left( \frac{a'}{a} \right)^2
 \right]\!,
\end{equation}
where $a' \equiv \partial a/\partial u$. Here, $a(u, \tau)$ is the
warped factor assumed to be $\tau$ dependent. For the shake of
simplicity we set $\psi_{1\mu}= \chi^i =0$ on the background
(\ref{DWans}) and then, the supersymmetry transformation
(\ref{susyvar1}) becomes
\begin{align}
\frac{1}{M_P} \delta\psi_{1u} &= D_u \,\epsilon_1 +
L(\tau)\gamma_u \epsilon^1 + \frac{\mathrm{i}}{2}Q_u(\tau)
\,\epsilon_1,
 \nonumber\\
\frac{1}{M_P} \delta\psi_{1\underline{\nu}} &=
\partial_{\underline{\nu}}\,\epsilon_1 +
\frac{1}{2}\gamma_{\underline{\nu}}\left(-\frac{a'}{a}\gamma_3
\epsilon_1 + {\mathrm{i}}{\rm e}^{K(\tau)/2M_P^2}\, W
\epsilon^1 \right) + \frac{\mathrm{i}}{2}
Q_{\underline{\nu}}(\tau) \,\epsilon_1
,\label{susyvar2}\\
\delta\chi^i &= {\mathrm{i}}\partial_\mu z^i \,
\gamma^\mu\epsilon^1 + N^i(\tau)\epsilon_1. \nonumber
\end{align}
Furthermore, all supersymmetric variations (\ref{susyvar2})
vanish  in order to have residual supersymmetry on the
ground states. We then simply assign
 that $\epsilon_1$ and $z^i$ depend on both $u$ and $\tau$.
Thus, the first equation in (\ref{susyvar2}) shows that
$\epsilon_1$ indeed depends  on $u$, while the second
equation gives a projection~equation
\begin{equation}\label{projector}
\frac{a'}{a}\gamma_3 \epsilon_1  = {\mathrm{i}}L(\tau)
\epsilon^1,
\end{equation}
which further gives
\begin{equation}\label{warp}
\frac{a'}{a}  = \pm \, \vert L(\tau) \vert.
\end{equation}
Thus, equation (\ref{warp}) shows that the warped factor
$a$ is indeed $\tau$-dependent, which is consistent with our
ansatz (\ref{DWans}). Introducing a real function
${\mathcal{W}}(\tau) \equiv  \lvert L(\tau) \rvert$, the
third equation in (\ref{susyvar2}) results in a set of BPS
equations
\begin{align}
z^i{'} &= \mp 2 g^{i\bar{j}}(\tau)
\bar{\partial}_{\bar{j}}{\mathcal{W}}(\tau),\nonumber\\
\bar{z}^{\bar{i}}{'} &= \mp 2 g^{j\bar{i}}(\tau)
\partial_j{\mathcal{W}}(\tau),
\label{gfe}
\end{align}
describing gradient flows. It is important to note that
using (\ref{gfe}) it can be shown that (\ref{warp}) is a
monotonic decreasing function and corresponds to the $c$
-function in the holographic correspondence \cite{DWc}.
Moreover, in the
 context of CFT the supersymmetric
 flows which is relevant for our analysis are described
 by a beta function
\begin{equation}\label{beta}
 \beta^i \equiv a \frac{dz^i}{da} = - 2g^{i \bar{j}}(\tau)
  \frac{\bar{\partial}_{\bar{j}}
 {\mathcal{W}}}{\mathcal{W}},
\end{equation}
 together with its complex conjugate, after employing (\ref{warp})
 and (\ref{gfe}). In this description the scalar fields play a role
 as coupling constants and the warp factor
 $a$ can be viewed as an energy scale~\cite{DWc,CDKV}.

In the analysis it is convenient that the potential
(\ref{V1}) can be shaped into
\begin{equation}\label{V11}
 {\mathcal{V}}(z,\bar{z}; \tau) = 4\, g^{i\bar{j}}\, \partial_i {\mathcal{W}}\,
 \bar{\partial}_{\bar{j}} {\mathcal{W}}
 - \frac{3}{M^2_P}\, {\mathcal{W}}^2,
\end{equation}
whose first derivative with respect to $(z,\bar{z})$ is given by a
set of the following equation
\begin{align}
\partial_i {\mathcal{V}} &= 4\, g^{j\bar{k}}\, \nabla_i
\partial_j {\mathcal{W}}\, \bar{\partial}_{\bar{k}} {\mathcal{W}}+
4\, g^{j\bar{k}}\, \partial_j {\mathcal{W}}\,
\partial_i \bar{\partial}_{\bar{k}} {\mathcal{W}}
 - \frac{6}{M^2_P}\, {\mathcal{W}}\,\partial_i{\mathcal{W}},\nonumber\\
\bar{\partial}_{\bar{i}}{\mathcal{V}} &= 4\, g^{j\bar{k}}\,
\bar{\nabla}_{\bar{i}}\bar{\partial}_{\bar{k}}{\mathcal{W}}\,
\partial_j{\mathcal{W}}+ 4\, g^{j\bar{k}}\,
\bar{\partial}_{\bar{k}}{\mathcal{W}}\, \bar{\partial}_{\bar{i}}
\partial_j {\mathcal{W}}
 - \frac{6}{M^2_P}\,
 {\mathcal{W}}\,\bar{\partial}_{\bar{i}}{\mathcal{W}},\label{dV}
\end{align}
where $\nabla_i\partial_j {\mathcal{W}}= \partial_i \partial_j
{\mathcal{W}} - \Gamma^k_{ij}\partial_k {\mathcal{W}}$. Note that
as mentioned in the preceding section, it is possible that the
theory becomes unphysical in the sense that it has wrong-sign
kinetic term for finite $\tau$. For example, it happens when the
flow described by (\ref{solKEM}) for $\Lambda
>0$ and $\tau > 1/2 \Lambda$. So, we have BPS domain walls whose
dynamics described by the warped factor $a(u,\tau)$ are
controlled by ghosts.

Now let us turn to consider the gradient flow equation
 (\ref{gfe}) and the first derivative of the scalar potential (\ref{dV}).\!\! Critical points of (\ref{gfe}) are~determined~by
\begin{equation}
\partial_i{\mathcal{W}}(\tau) = \bar{\partial}_{\bar{i}}{\mathcal{W}}(\tau)
=0,
\end{equation}
which leads to
\begin{equation}
\partial_i {\mathcal{V}} = \bar{\partial}_{\bar{i}} {\mathcal{V}} =0.
\end{equation}
It turns out that the critical points of
${\mathcal{W}}(\tau)$ are related to the Lorentz invariant
vacua described by the $N=1$ scalar potential
$V(z,\bar{z};\tau)$. Moreover,\vadjust{\pagebreak} the
existence of such points can be checked by the beta
function (\ref{beta}), which means that these could be in
the ultraviolet (UV) region if $a \to \infty$ or in the
infrared (IR) region if $a \to 0$. These aspects will be
addressed in Section~\ref{SUSYKEM} and
Section~\ref{SUSYKRS}.

\section{Supersymmetric vacua on K\"ahler--Einstein manifold}\label{SUSYKEM}

The aim of this section is to present analysis of
deformation of supersymmetric Lorentz invariant ground
states on the K\"ahler--Ricci flow when the initial
geometry is K\"ahler--Einstein geometry. As mentioned
above, these vacua correspond to critical points of
(\ref{gfe}). In particular, our interest is the AdS vacua
which correspond to the CFT. Our analysis here is in the
context of the Morse theory \cite{JM}\footnote{An excellent
background of Morse theory is also given, for example, in
\cite{YM}.} and the Morse--Bott theory \cite{BH,BH1}, and
in addition, applying the RG flow analysis to check the
existence of such vacua. We organize this section into two
parts. First, we perform the analysis to the vacua of the
global $N=1$ theory. Then we generalize it to the local
$N=1$ theory which are related to the BPS domain wall
solutions. We leave the proof of some theorems here in
Appendix~B.

\subsection{Global supersymmetric case}\label{GSC}

Let us first investigate some properties of supersymmetric
Lorentz invariant ground states of the theory where all
fermions and $\partial_{\mu} z^i$ are set to be
zero.\footnote{For the rest of the paper we refer Lorentz
invariant vacuum as vacuum or ground state.} In a smooth
region, \textit{i.e.} there is no collapsing flow, at the
ground state $p_0 \equiv (z_0, \bar{z}_0)$ we find that
supersymmetric conditions
\begin{equation}\label{susycon}
\partial_i W(z_0) = \bar{\partial}_{\bar{i}} \bar{W}(\bar{z}_0) =0,
\end{equation}
imply the set of equation in (\ref{dVz}) and supersymmetry
transformation (\ref{susyvar}) to be vanished. In other
words, critical point of holomorphic superpotential $W(z)$
defines a vacuum of the theory and $z_0$ does not depend on
$\tau$. Thus, the vacuum is fixed in this case. Moreover,
it is easy to see from (\ref{dVt}) that at the vacua the
scalar potential and the mass matrix of fermions do not
change with respect to $\tau$. So in order to characterize
the ground states, we have to consider the second-order
derivative of the scalar potential
(\ref{V})\vadjust{\pagebreak} with respect to $(z,\bar{z})$
called Hessian matrix evaluated at $p_0$, whose nonzero
component has the~form
\begin{equation}\label{HessV}
\partial_i \bar{\partial}_{\bar{j}} V(p_0, \tau)=
\sigma(\tau)^{-1} g^{l\bar{k}}(p_0 ;0) \partial_i
\partial_j W(z_0) \,
\bar{\partial}_{\bar{k}}\bar{\partial}_{\bar{j}}\bar{W}(\bar{z}_0),
\end{equation}
where $g_{l\bar{k}}(0)$ is a K\"ahler--Einstein metric and
we have assumed here that $g_{l\bar{k}}(p_0,0)$ is
invertible. Then (\ref{HessV}) leads to the following
statements:

\newtheorem{Vmorse}{Lemma}[section]

\begin{Vmorse}\label{LemmaVmorse}
The scalar potential $V(\tau)$ is a Morse function if and only if
the superpotential $W(z)$ has no degenerate critical points.
\end{Vmorse}

We want to mention here that the above lemma implies that
the superpotential is at least general quadratic function,
namely, $W(z) = a_{ij} (z-z_0)^i (z-z_0)^j$, where the
matrix coupling $(a_{ij})$ is invertible.\footnote{For
constant and linear superpotential, we have degenerate
supersymmetric vacua of the global theory in the sense that
they are trivial in the second-order analysis. On the other
side, all nonsupersymmetric vacua are degenerate in this
global case.} Correspondingly, all eigenvalues of the
fermion mass matrix (\ref{fermass}) evaluated at $p_0$ are
nonzero. For the rest of the paper, if $V(\tau)$ is a Morse
function, then we name it the Morse potential. Furthermore,
we have the following consequence.

\newtheorem{IsolV}[Vmorse]{Corollary}

\begin{IsolV}\label{Colisol}
All supersymmetric ground states of the Morse potential
$V(\tau)$ are nondegenerate $($isolated$)$.
\end{IsolV}

\noindent Therefore, since $V(\tau)$ is a Morse function we
can assign any supersymmetric vacuum by Morse index
describing the number of negative eigenvalue of the Hessian
matrix (\ref{HessV}).\footnote{These negative eigenvalues
describe the unstable directions of a vacuum.} Looking at
(\ref{HessV}), we find that for $\tau \ge 0$ and~$\Lambda >
0$
\begin{equation}\label{HessV0}
\partial_i \bar{\partial}_{\bar{j}} V(p_0, \tau > 1/2\Lambda)=
- \partial_i \bar{\partial}_{\bar{j}} V(p_0, \tau <
1/2\Lambda),
\end{equation}
which shows the existence of parity transformation of the
Hessian matrix at $p_0$ caused by the metric (\ref{solKEM})
mapping the Morse index from, say $\lambda$, to another
Morse index $2n_c - \lambda$. So, we can write our main
result of global supersymmetric vacua on K\"ahler--Einstein
manifold as follows.

\newtheorem{Modmorse}[Vmorse]{Theorem}

\begin{Modmorse}\label{Modmorselemma}
If  $p_0$ is an isolated supersymmetric vacuum of the Morse
index $\lambda,$ then there exist a parity transformation
of the matrix {\rm (\ref{HessV})} that changes the index
$\lambda$ to another Morse index $(2n_c -\lambda)$ due to
the deformation of K\"ahler--Einstein manifold.
Furthermore, around $p_0$ we can introduce real local
coordinates $X_p(\tau)= \vert \sigma(\tau) \vert^{-1/2}
X_p(0)$ with $p= 1,\ldots, 2n_c$ such~that
\begin{equation}\label{VM}
V(\tau) = \varepsilon(\sigma)\Big(- X_1^2(\tau)-\cdots-
X_{\lambda}^2(\tau) + X_{\lambda +1}^2(\tau) +\cdots+
X_{2n_c}^2(\tau) \Big),
\end{equation}
where
\begin{equation}\label{varep}
\varepsilon(\sigma) =\begin{cases}
\phantom{-}1 &   {\text{if}} \;\;\;  0 \le \tau < 1/2\Lambda, \\
-1 &  {\text{if}} \;\qquad \; \, \tau > 1/2\Lambda,\\
\end{cases}
\end{equation}
with $\Lambda >0$ for $\tau \ge 0$.
\end{Modmorse}
Note that for $\Lambda < 0$, the factor $\sigma(\tau)
\equiv (1- 2\Lambda \tau) > 0$, and in this case we have
only $\varepsilon(\sigma)=1$ for $\tau \ge 0$. Similar
result can also be obtained for static case,
 namely, for $\Lambda =0$ or the Calabi-Yau manifold.

Let us start to discuss Theorem~\ref{Modmorselemma} by
taking the example discussed in the previous section where
the initial geometry is a  K\"ahler--Einstein geometry
${\bf M}_0$ with positive definite metric and $\Lambda >
0$. Looking at (\ref{RiccKEM1}), we find that the flow
gives the same shape as the initial geometry ${\bf M}_0$
for $0 \le \tau < 1/2\Lambda$, while  for $\tau >
1/2\Lambda$ the flow turns into a K\"ahler--Einstein
geometry ${\bf \widehat{M}}_0$ with negative definite
metric  and $\widehat{\Lambda} < 0$. Therefore, this
geometrical deformation causes the parity transformation of
the matrix (\ref{HessV}) that changes the index $\lambda$
for $\tau < 1/2\Lambda$ to $(2n_c -\lambda)$ for $\tau >
1/2\Lambda$ of the same ground state.

\subsection{Local supersymmetric case}\label{LSC}

Similar as preceding subsection  we consider here the
smooth region where the geometric flow does not vanish. Our
interest here is to study  supersymmetric vacuum $p_0
\equiv (z_0, \bar{z}_0)$ in local case which demands~that
\begin{equation}\label{susycon1}
 \partial_i {\mathcal{W}}(p_0;\tau) =0  \Rightarrow \partial_i W(z_0) + \sigma(\tau)
 \frac{K_i(p_0; 0)}{M_P^2} W(z_0) =0,
\end{equation}
has to hold and correspondingly, both BPS equations in (\ref{gfe})
and the beta function (\ref{beta}) vanish. Then the
$\tau$-dependent scalar potential (\ref{V1}) becomes
\begin{equation}\label{VAdS}
{\mathcal{V}}(p_0; \tau) =  - \frac{3}{M_P^2} {\rm
e}^{\sigma(\tau)K(p_0; 0)/M_P^2}\vert W(z_0)\vert^2 =
-\frac{3}{M_P^2}{\mathcal{W}}^2(p_0;\tau),
\end{equation}
which is zero or negative definite standing for
cosmological constant of the spacetime on which the Ricci
scalar (\ref{Rans}) has the~form
\begin{equation}\label{Rans1}
R =  12 \, {\mathcal{W}}^2(p_0;\tau),
\end{equation}
since $(a'/a)'(p_0) =0$.  Hence, the warped factor $a(u, \tau)$ is
simply
\begin{equation}\label{warp1}
a(u, \tau) =  a_0(\tau) \, {\mathrm{exp}}\Big\lbrack \pm
{\mathcal{W}}(p_0;\tau)u \Big\rbrack.
\end{equation}
Furthermore, the Hessian matrix of the scalar potential
(\ref{V1}) in this case is given~by
\begin{align}
\partial_i \partial_j {\mathcal{V}}(p_0; \tau)  &=- \frac{1}{M^2_P}
{\mathcal{M}}_{ij}(p_0; \tau) \bar{L}(p_0; \tau), \nonumber\\[4pt]
\bar{\partial}_{\bar{i}} \bar{\partial}_{\bar{j}}
{\mathcal{V}}(p_0; \tau) &=- \frac{1}{M^2_P}
\bar{{\mathcal{M}}}_{\bar{i}\bar{j}}(p_0;
\tau) L(p_0; \tau), \label{HessV1}\\[4pt]
 \partial_i \bar{\partial}_{\bar{j}}{\mathcal{V}}(p_0; \tau) &=
 \sigma(\tau)^{-1}g^{k\bar{l}}(p_0; 0) {\mathcal{M}}_{ik}(p_0; \tau)
 \bar{{\mathcal{M}}}_{\bar{l}\bar{j}}(p_0; \tau)\notag\\[4pt]
  &\quad - \frac{2\sigma(\tau)}{M^4_P} g_{i\bar{j}}(p_0; 0)
  {\mathcal{W}}^2(p_0;\tau) , \nonumber
\end{align}
where
{\fontsize{10.40}{10.95}\selectfont\begin{align}\label{fermass2}
{\mathcal{M}}_{ij}(p_0; \tau)&= e^{\sigma(\tau)K(p_0;
0)/2M_P^2}\notag\\[4pt]
&\quad\times\left(\!\partial_i  \partial_j W(z_0)\,{+}\,
\frac{\sigma(\tau) }{M_P^2}K_{ij}(p_0; 0)W(z_0)\,{+}\,
\frac{\sigma(\tau) }{M_P^2} K_j(p_0; 0)\partial_i
W(z_0)\!\right)\!,\vspace*{6pt}
\end{align}}
\noindent and we have assumed that $g_{l\bar{k}}(p_0,0)$ is
invertible. As mentioned in the previous subsection, the
Morse index of a vacua is given by the negative eigenvalues
of (\ref{HessV1}) describing unstable directions. Along
these directions, the gradient flows provided by
(\ref{gfe}) are unstable and therefore, we have unstable
walls in the context of dynamical system. On the other
side, we obtain a stable solution along stable direction
described by the positive eigenvalues of (\ref{HessV1}).
Finally, to check the existence of such vacua in the IR and
UV regions, we need to consider a first-order expansion of
the beta function (\ref{beta}) at $p_0$ provided~by
\begin{equation}\label{beta1}
{\mathcal{U}} \equiv - \left(
\begin{matrix}
  \partial_j \beta^i  & & \partial_j \bar{\beta}^{\bar{i}} \\[4pt]
  \bar{\partial}_{\bar{j}} \beta^i  & &
  \bar{\partial}_{\bar{j}} \bar{\beta}^{\bar{i}} \\
\end{matrix}
\right)(p_0; \tau),
\end{equation}
where
\begin{align}
\partial_j \beta^i(p_0; \tau) &= - 2
\sigma(\tau)^{-1}g^{i \bar{k}}(p_0; 0) \frac{\partial_j
\bar{\partial}_{\bar{k}}
 {\mathcal{W}}(p_0; \tau)}{{\mathcal{W}}(p_0; \tau)} =
 -\frac{1}{M_P^2}\, \delta^i_j,\nonumber\\[4pt]
\bar{\partial}_{\bar{j}} \beta^i(p_0; \tau) &= - 2
\sigma(\tau)^{-1}g^{i \bar{k}}(p_0; 0)
\frac{\bar{\partial}_{\bar{j}} \bar{\partial}_{\bar{k}}
 {\mathcal{W}}(p_0; \tau)}{{\mathcal{W}}(p_0; \tau)},
\end{align}
and the rests are their complex conjugate. In order to have
a vacuum in the UV region, at least one eigenvalue of the
matrix (\ref{beta1}) must be positive, because the RG flow
departs the region in this direction. On the
other\vadjust{\pagebreak} side, in the IR region the RG
flow approaches a vacuum in the direction of negative
eigenvalue of (\ref{beta1}). In the following, we consider
two cases, namely the Morse theory analysis for
nondegenerate vacua and then, degenerate vacua using
Morse--Bott theory.

Let us first discuss the case in which the scalar potential
(\ref{V1}) is Morse potential. For $W(z_0) =0$, the ground
states are flat Minkowskian spacetime and we regain the
condition (\ref{susycon}), which means that $p_0$ is fixed
and does not evolve with respect to $\tau$. Additionally,
the warped factor equals $a_0(\tau)$ and can be rescaled by
coordinate redefinition. In this case, the only nonzero
coupling is the fermionic mass matrix defined in
(\ref{fermass1}) which deforms with respect to $\tau$ as
one can directly see from (\ref{fermass2}). This follows
that the non zero components of the Hessian
(\ref{HessV1})~are
\begin{equation}
 \partial_i \bar{\partial}_{\bar{j}}{\mathcal{V}}(p_0; \tau)
 = {\rm e}^{\sigma(\tau)K(p_0; 0)/M_P^2}
\sigma(\tau)^{-1} g^{l\bar{k}}(p_0 ;0)\, \partial_i
\partial_l W(z_0) \,
\bar{\partial}_{\bar{k}}\bar{\partial}_{\bar{j}}\bar{W}(\bar{z}_0),
\end{equation}
which means that the nondegeneracy of the vacua is
determined by the holomorphic superpotential $W(z)$.
Moreover, the property of the vacua does not change if we
set $M_P \to +\infty$, so the analysis is similar to the
global case. In other words, they have the same properties
as in the second-order analysis for $W(z_0)=0$. Then, in
this case the holomorphic superpotential $W(z)$ has at
least general quadratic form discussed in the preceding
subsection. Note that in this case the matrix (\ref{beta1})
is meaningless because these are not related to the CFT in
three dimensions and, additionally, it diverges.

The other case, namely $W(z_0) \ne 0 $, is curved symmetric
spacetime with negative cosmological constant, called AdS.
In this case, the vacuum $p_0$ depends on $\tau$ if
$K_j(p_0; 0) \ne 0$, which means that our ground state
$p_0(\tau)$ and, moreover, all couplings
(\ref{gravitinomass}) to (\ref{V1}) do deform with respect
to $\tau$.\footnote{Since the vacua depend on $\tau$, so
the metric (\ref{solKEM}) may change the nature of the
vacua. Therefore, such vacua are dynamical with respect to
$\tau$.} In addition, we assume that the sign of the
initial metric evaluated at $p_0(\tau)$, namely
$g_{i\bar{j}}(p_0(\tau) ;0)$, is unchanged and well defined
for $\tau \ge 0$ but not at the singular flow. Then from
(\ref{warp1}) we can choose, for example, a model with
positive sign where for finite $\tau$, the UV region can be
defined around $u \to +\infty$ and $a \to +\infty$.
Additionally, the IR region in this case is around $u \to
-\infty$ and~$a \to 0$.

Unlike Minkowskian cases, in AdS vacua we cannot trivially
observe geometrical change of the vacua after the metric
(\ref{solKEM}) hits the singularity at $\tau = 1/2\Lambda$
with $\Lambda >0$ indicated by the parity transformation of
the Hessian matrix (\ref{HessV1}). Therefore, one has to
enforce additional conditions in order to achieve such
particular situation. The next step is to analyze the
eigenvalues of the matrix
(\ref{beta1}) whether such vacua exist in the UV and IR regions.

Before coming to the result described by the theorem below,
let us define some quantities related to the Hessian matrix
(\ref{HessV1}) in the following:
\begin{align}
{\mathcal{V}}_{ij}(p_0(\tau); \tau)  & \equiv
 - \frac{\varepsilon(\sigma)}{M^2_P} \, {\mathcal{M}}_{ij}(p_0(\tau); \tau)
 \bar{L}(p_0(\tau); \tau),\label{Vij}\\
 {\mathcal{V}}_{i\bar{j}}(p_0(\tau);\tau) & \equiv \vert \sigma(\tau)\vert^{-1}
 g^{k\bar{l}}(p_0(\tau); 0) {\mathcal{M}}_{ik}(p_0(\tau); \tau)
 \bar{{\mathcal{M}}}_{\bar{l}\bar{j}}(p_0(\tau); \tau) \nonumber\\
   &\quad- \frac{2 \vert \sigma(\tau) \vert}{M^4_P} g_{i\bar{j}}(p_0(\tau); 0)
  {\mathcal{W}}^2(p_0(\tau); \tau), \nonumber
\end{align}
where $\varepsilon(\sigma)$ is given in (\ref{varep}).
Particular result in a nondegenerate AdS case where the
parity transformation of the Hessian matrix (\ref{HessV1})
is manifest can then be written as follows.

\newtheorem{Vmorse1}[Vmorse]{Theorem}

\begin{Vmorse1}\label{LemmaVmorse1}
Let ${\mathcal{V}}(\tau)$ be a Morse potential and
$p_0(\tau) = q_0$ be an isolated AdS ground state of the
Morse index $\lambda$ for $0 \le \tau < 1/2\Lambda$ with
$\Lambda >0$. Then, there are real local coordinates
$Y_p(\tau)$ around $q_0$ with $p = 1,\ldots,2n_c$ such~that
\begin{equation}
{\mathcal{V}}(\tau) = {\mathcal{V}}(q_0; \tau) -
Y_1^2(\tau) -\cdots- Y_{\lambda}^2(\tau)
 + Y_{\lambda +1}^2(\tau) +\cdots+ Y_{2n_c}^2(\tau).
\end{equation}
Suppose that for all $i,j = 1,\ldots,n_c$ the following
inequalities
\begin{align}
{\mathrm{Re}}\big({\mathcal{V}}_{ij}(p_0(\tau); \tau) \big)
> 0, &\quad {\mathrm{Im}}\big({\mathcal{V}}_{ij}(p_0(\tau);
\tau)
\big)> 0, \nonumber\\
{\mathrm{Re}}\big({\mathcal{V}}_{i\bar{j}}(p_0(\tau);
\tau)\big) > 0, &\quad
{\mathrm{Im}}\big({\mathcal{V}}_{i\bar{j}}(p_0(\tau);
\tau)\big) > 0,  \label{mirrorcon}
\end{align}
hold for $\tau \ge 0$ and $\tau \ne 1/2\Lambda$.
 Let $p_0(\tau) = \hat{q}_0$ be another isolated vacuum for $\tau >
1/2\Lambda $ and both $(q_0, \hat{q}_0)$ exist in the UV or
IR regions. Then, K\"ahler--Ricci flow causes a parity
transformation of the Hessian matrix {\rm (\ref{HessV1})},
that maps $q_0$ of the index $\lambda$ to $\hat{q}_0$ of
the index $(2n_c -\lambda)$ such that around $\hat{q}_0$
real local coordinates $\widehat{Y}_p(\tau) \ne Y_p(\tau)$
exist~and
\begin{equation}
{\mathcal{V}}(\tau) = {\mathcal{V}}(\hat{q}_0; \tau) +
\widehat{Y}_1^2(\tau) +\cdots+
\widehat{Y}_{\lambda}^2(\tau)
 - \widehat{Y}_{\lambda +1}^2(\tau) -\cdots- \widehat{Y}_{2n_c}^2(\tau),
\end{equation}
for $\tau > 1/2\Lambda$.
\end{Vmorse1}

Our comments of the above theorem are in order. Firstly, in
general $q_0 \ne \hat{q}_0$ and they are called parity pair
of the vacua in the sense that $q_0$ has $\lambda$ negative
eigenvalues of the Hessian matrix (\ref{HessV1}), while
$2n_c - \lambda$ negative eigenvalues belong to $\hat{q}_0$
caused by the geometric flow (\ref{solKEM}) after passing
the singularity at $\tau = 1/2\Lambda$. On the other side,
if the inequalities (\ref{mirrorcon}) do not hold, then
$q_0$ and $\hat{q}_0$ would have the same index~$\lambda$.

Secondly, we can reconsider the case discussed in the
global case where the initial geometry is taken to be a
K\"ahler--Einstein geometry ${\bf M}_0$ with positive
definite metric and $\Lambda > 0$. However, in the
dynamical AdS case since $p_0(\tau)$, the holomorphic
superpotential $W(z_0(\tau))$ together with the fermion
mass matrix (\ref{fermass2}) also control the properties of
the ground states in the second-order analysis described by
(\ref{HessV1}). In other words, there are two aspects that
play a role in the second-order analysis of vacua, namely
the K\"ahler--Ricci flow (\ref{solKEM}) and the dynamics of
coupling quantities which depend both on the form of the
superpotential and the geometry. Note that
$g_{i\bar{j}}(p_0(\tau); 0)$ remains positive definite for
$\tau \ge 0$ in this~case.

Thirdly, the inequalities (\ref{mirrorcon}) mean that
 all components of the matrix (\ref{HessV1}) have at least to
 change the sign as the K\"ahler--Einstein
 geometry evolves. This can be easily seen in a special case where
\begin{equation}\label{mirrorcon1}
{\mathcal{M}}_{ij}(p_0(\tau); \tau)  =0,
\end{equation}
 holds for $\tau \ge 0$ and $\tau \ne 1/2\Lambda$. Consequently,
 we have a theory in which
 all spin-$\frac{1}{2}$ fermionic are massless. Then
the Hessian matrix (\ref{HessV1}) has simple diagonal~form
\begin{equation}
\partial_i \bar{\partial}_{\bar{j}}{\mathcal{V}}(p_0(\tau); \tau) =
    - \frac{2\sigma(\tau)}{M^4_P} \, g_{i\bar{j}}(p_0(\tau); 0)
 \, {\mathcal{W}}^2(p_0(\tau); \tau).
\end{equation}
Hence, in this case, the condition (\ref{mirrorcon}) can be
simplified by stating that the sign of
$g_{i\bar{j}}(p_0(\tau); 0)$ is fixed with respect to
$\tau$ in order the parity to be revealed. In $\Lambda >0$
case, we have Morse index $2n_c$ for $\tau < 1/2\Lambda$
and then, it changes to $0$ for $\tau > 1/2\Lambda$. In
other words, we have unstable walls for $\tau <
1/2\Lambda$, which become stable after hitting the
singularity at $\tau = 1/2\Lambda$. Such situation,
however, does not occur in $\Lambda <0$ case and the index
is still $2n_c$ for~$\tau \ge 0$.

Furthermore, the condition (\ref{mirrorcon1}) also leads to
\begin{equation}\label{mirrorcon2}
\partial_i \partial_j {\mathcal{W}}(p_0(\tau); \tau)  =0.
\end{equation}
Then, the matrix (\ref{beta1}) becomes simply
\begin{equation}
{\mathcal{U}} =  \frac{1}{M_P^2} \,\left(
\begin{matrix}
   \delta^i_j & & 0 \\
  & & \\
  0 & & \delta^{\bar{i}}_{\bar{j}}
   \\
\end{matrix}\right), \label{beta2}
\end{equation}
which shows that these vacua only live in the UV region.

On the other hand, it is also of interest to consider a
case where the ground states are determined by the critical
points of the holomorphic superpotential $W(z)$, i.e.,
equation (\ref{susycon}) is fulfilled. Then the $U(1)$-connection (\ref{U1conKE}) evaluated at $p_0$ vanishes in
all order analysis.\footnote{This case has been considered
in reference \cite{GZA} for model with a chiral
multiplet.} In other words, the supersymmetric condition
(\ref{susycon1}) reduces to
\begin{equation}
\partial_i W(z_0) = 0, \quad W(z_0) \ne 0
,\label{susycon2}
\end{equation}
In this case, the ground state $p_0$ is coming from the
critical points of $W(z)$ fixed with respect to $\tau$. So,
the second-order properties of vacua described by
(\ref{HessV1}) are fully controlled by the geometric flow
and the parity transformation appears if the condition
(\ref{mirrorcon}) is also fulfilled.

We finally put some remarks for the case of Morse
potential. In dynamical AdS vacua if the condition
(\ref{mirrorcon}) is not satisfied, then the parity  does
not emerge but the deformation of the vacua still exists
due to the K\"ahler--Ricci flow. These vacua admit the same
Morse index $\lambda$, see the above example for $\Lambda
<0$ and $\tau \ge 0$. Similar things happen for the static
case. In addition, since the scalar potential (\ref{V1}) is
a Morse potential, then it is impossible to have a
degenerate vacuum. Therefore, any vacuum does not deform to
another vacuum with different Morse index in $0 \le \tau <
1/2\Lambda$ with $\Lambda >0$. Thus, no index modification
of the vacua exists in the interval.

Now we turn to consider when the ground states are
degenerate.
 This can occur if the Hessian matrix (\ref{HessV1})
  has $m$ zero eigenvalues
with $m \le 2n_c$. Therefore, the vacua can be viewed as
$m$-dimensional submanifold of the K\"ahler--Einstein
manifold. For the case at hand, the scalar potential
(\ref{V1}) is a Morse--Bott function (or Morse--Bott
potential), which is a generalization of a Morse function.
If the vacua are Minkowskian, then the superpotential
$W(z)$ must be constant or take a linear form as discussed
in Section \ref{GSC} since the other forms are excluded.
Thus, we have $m =2n_c$ and the vacuum manifold could then
be the K\"ahler--Einstein manifold ${\bf M}_0$ or ${\bf
\widehat{M}}_0$ depending on $\tau$ and no parity
transformation of the matrix (\ref{HessV1}) caused by the
flow (\ref{solKEM})
 appears in this model. Note that in the model
 at hand we also have singular (\ref{beta1}) because it is undefined
 for $\partial_i \partial_j {\mathcal{W}}(p_0(\tau); \tau)
 = {\mathcal{W}}(p_0(\tau); \tau) =0$. Hence, this means that the
 three-dimensional CFT does not exist and we do not have the correspondence
 in the Minkowskian ground states.

In AdS vacua our construction is as follows. Let $S$ be an
$m$-dimensional vacuum submanifold of a K\"ahler--Einstein
geometry ${\bf M}_0$.\footnote{We could also consider ${\bf
M}_0$ as a set of a disjoint union of some connected smooth
manifolds with finite dimension, see, for example,
\cite{BH,BH1}.} Then at any $p_0 \in S$ we can split the
tangent space $T_{p_0}{\bf M}_0$~as
\begin{equation}
T_{p_0}{\bf M}_0 = T_{p_0}S \oplus N_{p_0}S,
\end{equation}
where $T_{p_0}S$ is the tangent space of $S$ and $N_{p_0}S$
is the normal space of $S$. Moreover, the Hessian matrix
(\ref{HessV1}) is nondegenerate in the normal direction
to~$S$.

Unlike the previous case, since the scalar potential is
Morse--Bott potential, thus we have a rich and complicated
structure of vacua, particularly in the dynamical case. In
the following we list some possibilities. The first
possibility is a similar situation as in Theorem
\ref{LemmaVmorse1} which can be written down in the
following statements.

\newtheorem{Vmorse2}[Vmorse]{Theorem}

\begin{Vmorse2}\label{LemmaVmorse2}
Let ${\mathcal{V}}(\tau)$ be Morse--Bott potential. At any
degenerate AdS vacuum $q_0 \in S$ where $S$ is an
$m$-dimensional vacuum submanifold of a\break
\hbox{K\"ahler--Einstein} geometry $\;{\bf M}_0$, then, in
the direction of $N_{q_0}S$ we can write the scalar
potential {\rm (\ref{V1})}~as
\begin{equation}
{\mathcal{V}}(\tau) = {\mathcal{V}}(S; \tau) -
Y_1^2(\tau)-\cdots- Y_{\underline{\lambda}}^2(\tau)
 + Y_{\underline{\lambda} +1}^2(\tau)+\cdots+ Y_{2n_c -m}^2(\tau),
\end{equation}
for $0 \le \tau < 1/2\Lambda$ with $\Lambda >0$. Suppose
that there exists another $m$-\break dimensional vacuum
submanifold $\widehat{S}$ of
 a K\"ahler--Einstein geometry $\;{\bf \widehat{M}}_0$ for
 $\tau > 1/2\Lambda $, the
inequalities {\rm (\ref{mirrorcon})} are satisfied, and
both submanifolds $(S, \widehat{S})$ live in the UV or IR
regions. Therefore, the deformation of $\;{\bf M}_0$ via
K\"ahler--Ricci flow causes a parity transformation of the
Hessian matrix {\rm (\ref{HessV1})} that changes $q_0 \in
S$  to $\hat{q}_0 \in \widehat{S}$  such that in the
direction of $N_{\hat{q}_0}\widehat{S}$ the scalar
potential {\rm (\ref{V1})} has the~form
\begin{equation}
{\mathcal{V}}(\tau) = {\mathcal{V}}(\widehat{S}; \tau) +
\widehat{Y}_1^2(\tau) +\cdots+
\widehat{Y}_{\underline{\lambda}}^2(\tau)
 - \widehat{Y}_{\underline{\lambda} +1}^2(\tau)-\cdots-
 \widehat{Y}_{2n_c -m}^2(\tau),
\end{equation}
 for $\tau > 1/2\Lambda$ with $\widehat{Y}_{\underline{p}}(\tau) \ne
Y_{\underline{p}}(\tau)$, where $\underline{p}
=1,\ldots,2n_c -m$.
\end{Vmorse2}
This theorem implies that each $q_0 \in S \subseteq {\bf
M}_0$ has the same index $\underline{\lambda}$ for $0 \le
\tau < 1/2\Lambda$. In other words, $S$ is a vacuum
submanifold of the index $\underline{\lambda}$. Moreover,
since the ground states are $\tau$-dependent, i.e.,
$p_0(\tau)$, and if the Hessian matrix (\ref{HessV1}) is
changed in sign after passing the singular point, then the
submanifold $S$ deforms to the submanifold $\widehat{S}
\subseteq {\bf \widehat{M}}_0$ of the index $(2n_c
-\underline{\lambda})$ for $\tau > 1/2\Lambda$. Both $S$
and $\widehat{S}$ are called parity pair for similar reason
as in the nondegenerate case, namely $\widehat{S}$ of the
index $(2n_c -\underline{\lambda})$ has the same dimension
as $S$ and exists after the geometric flow (\ref{solKEM})
hits the singularity which splits ${\bf M}_0$ and ${\bf
\widehat{M}}_0$.

On the other side, in the static case where the
$U(1)$-connection evaluated at $p_0$ is zero, the vacuum
submanifold $S$ stays fixed, which means that it does not
change with respect to $\tau$. In other words, we have $S =
\widehat{S}$, but its index could change due to
K\"ahler--Ricci~flow.

The last possible situation is as follows.

\newtheorem{Vmorse3}[Vmorse]{Theorem}

\begin{Vmorse3}\label{LemmaVmorse3}
 Suppose ${\mathcal{V}}(\tau)$ is a Morse--Bott potential and the
 conditions {\rm (\ref{mirrorcon})} do not hold. Let $S$ be an $m$-dimensional
 vacuum submanifold of ${\bf M}_0$ with index $\underline{\lambda}$
 in $0 \le \tau < \tau_0$ and
 $\tau_0 < 1/2\Lambda$. Then we have the following cases.\break
 $S$ deforms~to
\begin{enumerate}
\item[{\rm 1.}]  an $n$-dimensional vacuum submanifold $S_1 \subseteq {\bf
M}_0$ of the index $\underline{\lambda}_1$ in $\tau_0 \le
\tau < 1/2\Lambda$. If $n \ne m$, then the index
$\underline{\lambda}_1 \in \{0,\ldots,2n_c -n \}$. However,
if $n =m$, then $\underline{\lambda}_1 \ne
\underline{\lambda}$ and $\underline{\lambda}_1 \in
\{0,\ldots,2n_c -n \}$.

 \item[{\rm 2.}] an $n_1$-dimensional vacuum submanifold
$\widehat{S}_1 \subseteq {\bf \widehat{M}}_0 $ of the index
$\underline{\lambda}_2$ in $\tau
> 1/2\Lambda$.  If $n_1 \ne m$, then we have
$\underline{\lambda}_2 \in \{0,\ldots,2n_c -n_1 \}$. But,
if $n_1 =m$, then $\underline{\lambda}_2 \ne 2n_c
-\underline{\lambda}$ and $\underline{\lambda}_2 \in
\{0,\ldots,2n_c -n_1 \}$.
\end{enumerate}
Furthermore, $S_1$ and $\widehat{S}_1$ are not the parity pair.
All $S$, $S_1$, and $\widehat{S}_1$ may exist in the UV or IR
regions.
\end{Vmorse3}
 This leads to a result that both the dimension and the index of
 the vacuum manifold may also be changed before or after the
 singularity, which is not related to the case stated in
 Theorem \ref{LemmaVmorse2}. For
example, there is a situation where a nondegenerate vacuum
changes to another degenerate vacuum as $\tau$ runs before or
after singularity. We give an example of this situation, namely
$\;\lC P^1$ model with linear superpotential, in the next
subsection.

\subsection{A model with linear superpotential}\label{MLS}

In this subsection, we discuss a $\;\lC {\rm P}^{n_c}$ model with
linear superpotential in order to make the statements in
the previous subsection clearer. For the model at hand, the
second-order derivative of the superpotential equals zero,
so some difficulties have been reduced.

Let us start by considering a superpotential $W(z)$ which
takes linear form~as
\begin{equation}
 W(z) = a_0 + a_i\,z^i,
\end{equation}
where the constants $a_0, a_i \in \lR$. Moreover, the
scalar manifold is chosen to be $\;\lC {\rm P}^{n_c}$ whose
K\"ahler potential has the~form
\begin{equation}
 K(z, \bar{z}; 0) = {\mathrm{ln}}(1 + z^i \bar{z}^{\bar{i}}),
\end{equation}
where $z^i$ is the standard Fubini--Study coordinates of
$\;\lC {\rm P}^{n_c}$. In the model, the metric is positive
definite with the constant $\Lambda = n_c + 1 >0$. Then the
K\"ahler--Ricci flow implies that the $\tau$-dependent
K\"ahler potential is given~by
\begin{equation}
 K(z, \bar{z}; \tau) = \sigma(\tau)
 \, {\mathrm{ln}}(1 + z^i \bar{z}^{\bar{i}} ),
\end{equation}
and the modified  constant is $\widehat{\Lambda}(\tau) \equiv (n_c
+ 1) / \sigma(\tau)$. So, in the interval $0 \le \tau < 1/2 (n_c
+1)$ the flow is diffeomorphic to $\;\lC {\rm P}^{n_c}$, but for  $ \tau
> 1/2(n_c +1)$ it becomes $\;\widehat{\lC {\rm P}^{n_c}}$ on which the resulting
metric is negative definite with the
constant~$\widehat{\Lambda} <0$.

As previously mentioned, this model has degenerate ground
states in the global supersymmetric theory, but in a local
supersymmetry it is still nontrivial. In local case, the
supersymmetric condition (\ref{susycon1}) is given~by
\begin{equation}\label{modvacon}
 a_i + \frac{ \sigma(\tau) \,\bar{z}^{\bar{i}}}{M_P^2(1 +
 z^k \bar{z}^{\bar{k}} )}(a_0 + a_j\,z^j) =0.
\end{equation}
It is easy to see that for the static case, $z^i =0$ is a
vacuum by demanding $a_i =0$. Furthermore, the Hessian
matrix (\ref{HessV1}) becomes simply diagonal matrix whose
nonzero elements are
\begin{equation}
\partial_i \bar{\partial}_{\bar{j}}{\mathcal{V}}(p_0; \tau) =
    - \frac{2\sigma(\tau)}{M^4_P} \, \delta_{i\bar{j}}
 \, \vert a_0 \vert^2.
\end{equation}
Thus we have an isolated AdS spacetime for $a_0 \ne 0$
representing unstable walls whose Morse index is $2n_c$ for $0\le
\tau < 1/2(n_c + 1)$. Then it changes to $0$ for $ \tau
> 1/2(n_c + 1)$ describing stable walls as the geometry evolves
 with respect to the K\"ahler--Ricci equation. Such vacua exist
 in the UV region since we have (\ref{beta2}). On the other hand,
 the ground states turn out to be degenerate Minkowskian spacetime
 if $a_0 =0$ for any
$z^i$. Therefore, the vacuum manifold is $\;\lC {\rm P}^{n_c}$ for $0
\le \tau < 1/2 (n_c +1)$ and then becomes $\;\widehat{\lC
{\rm P}^{n_c}}$ for  $ \tau
> 1/2(n_c +1)$.

In nonstatic case, it is convenient to take $n_c =1$ case.
So the ground state is chosen to~be
\begin{equation}
z_0 = (x_0 + {\mathrm{i}}y_0),
\end{equation}
where $x_0, y_0 \in \lR$. In this case, the condition
(\ref{modvacon}) results~in
\begin{align}
x_0(\tau) &= \frac{1}{2(\sigma(\tau)+M_P^2)}\!\left\lbrack
{-}\frac{\sigma(\tau) a_0 }{a_1}
\pm\!\left(\!\left(\frac{\sigma(\tau) a_0
}{a_1}\right)^2\,{-}\,4 M_P^2
(\sigma(\tau)+M_P^2) \right)^{\!\!1/2} \right\rbrack\!, \nonumber\\
 y_0(\tau) &= 0, \label{vacuum}
\end{align}
with $a_1 \ne 0$. Then, in order to obtain a well-defined
theory, the parameter $\tau$ must fulfill the following
inequalities,
\begin{align}
\tau & \le \frac{1}{4} - \frac{M_P^2}{2} \left(\frac{a_1
}{a_0}\right)^2
 \left\lbrack
1 + \left( 1 + \left( \frac{a_0}{a_1}\right)^2 \right)^{1/2}
\right\rbrack\!, \nonumber\\
\tau & \ge \frac{1}{4} + \frac{M_P^2}{2} \left(\frac{a_1
}{a_0}\right)^2 \left\lbrack -1 + \left(1 + \left(
\frac{a_0}{a_1}\right)^2 \right)^{1/2} \right\rbrack\!,
\label{welldef}
\end{align}
before and after singular point $\tau = 1/4$. The
eigenvalues of the Hessian matrix (\ref{HessV1}) in this
case have the~form
\begin{align}
\lambda^{\mathcal{V}}_{1,2}(\tau)&= \frac{2
\lvert\sigma(\tau)\rvert}{M_P^4}\left( 1 + x^2_0
\right)^{-2+ \sigma(\tau)/M_P^2}  \bigg({-}2
\varepsilon(\sigma) \left\lbrack\vphantom{\frac{1}{2}} (a_0 + a_1 x_0)^2\right.\nonumber\\
&\quad-\left.\frac{1}{2} (a_1 - a_0 x_0)^2 \, x_0^2
\right\rbrack\pm\big\vert(a_0 + a_1 x_0)(a_1 - a_0 x_0)x_0
\big\vert \bigg), \label{eigenV}
\end{align}
where $\varepsilon(\sigma)$ is given by (\ref{varep}) with
$\Lambda =2$. Moreover, the existence of a vacuum can be
checked by (\ref{beta1}) whose eigenvalues~are
\begin{equation}\label{eigenU}
\lambda^{\mathcal{U}}_{1,2}(\tau) = \frac{1}{M_P^2} \left( 1
 \pm  \,  \left\vert \frac{ x_0 (a_1 - a_0 x_0)}
 {(a_0 + a_1 x_0)}\right\vert \, \right).
\end{equation}
Let us simplify the model in which both $a_0$ and $a_1$ are
positive and $a_0 \gg a_1$. For $\tau \lesssim \frac{1}{4}
- \frac{a_1 M_P^2}{2a_0}$ we have $\varepsilon(\sigma) =1$,
and then local minimum occurs~in
\begin{equation}\label{locmin}
  - \frac{a_0 M_P^2}{8 a_1 \sqrt{2}} < \tau \lesssim \frac{1}{4}
  - \frac{a_1 M_P^2}{2a_0},
\end{equation}
while local maximum exists in
\begin{equation}\label{locmax}
 \tau <  - \frac{a_0 M_P^2}{8 a_1}.
\end{equation}
Additionally, saddle arises between
\begin{equation}\label{saddle}
  - \frac{a_0
M_P^2}{8 a_1}  < \tau <   - \frac{a_0 M_P^2}{8 a_1
\sqrt{2}}.
\end{equation}
Also, it is possible to have intrinsic degenerate
 vacua\footnote{Intrinsic means that these degenerate vacua
 are coming from degenerate critical points of ${\mathcal{W}}$
  \cite{GZA}.} if
\begin{equation}\label{degvac}
(a_0 + a_1 x_0)^2 = (a_1 - a_0 x_0)^2 \, x_0^2
\end{equation}
holds, which occurs around
\begin{equation}\label{intrtau}
\tau \approx - \frac{a_0 M_P^2}{8 a_1},
\end{equation}
whereas the other degeneracy needs
\begin{equation}\label{degvac1}
4(a_0 + a_1 x_0)^2 = (a_1 - a_0 x_0)^2 \, x_0^2
\end{equation}
and takes place near
\begin{equation}
\tau \approx  - \frac{a_0 M_P^2}{8 a_1 \sqrt{2}}.
\label{degtau}
\end{equation}
These mean that a nondegenerate ground state can be changed
into another degenerate ground state by K\"ahler--Ricci
flow far before singularity at $\tau = 1/4$. In other
words, this situation occurs without geometrical change and
further shows that the scalar potential of the model is
Morse--Bott potential. Then, the analysis using
(\ref{eigenU}) shows that all vacua may exist in the UV
region because it has at least one positive eigenvalue that
ensures the RG flows depart the region. However, in the IR
region, local maximum and intrinsic degenerate vacua do not
exist and therefore, no RG flows approach the vacua.

In  $\tau \gtrsim \frac{1}{4} + \frac{a_1 M_P^2}{2a_0}$ we
get $\varepsilon(\sigma) = -1$, which leads to the fact
that this model admits parity transformation of the Hessian
matrix (\ref{HessV1}). Then, the following interval
\begin{equation}\label{locmax1}
\frac{1}{4} + \frac{a_1 M_P^2}{2a_0} \lesssim \tau <
\frac{a_0 M_P^2}{8 a_1 \sqrt{2}}
\end{equation}
 shows the existence of local maximum, whereas
\begin{equation}\label{locmin1}
 \tau >  \frac{a_0
M_P^2}{8 a_1}
\end{equation}
 is for local minimum.  Next, saddle exists for
\begin{equation}\label{saddle1}
 \frac{a_0 M_P^2}{8 a_1 \sqrt{2}}   < \tau <
 \frac{a_0 M_P^2}{8 a_1},
\end{equation}
 while intrinsic degenerate vacua occur near
\begin{equation}\label{intrtau1}
\tau \approx  \frac{a_0 M_P^2}{8 a_1},
\end{equation}
and the other degenerate vacua arise around
\begin{equation}\label{degtau1}
\tau \approx  \frac{a_0 M_P^2}{8 a_1 \sqrt{2}}.
\end{equation}
 Similar as in the previous case, in the UV region we have all
possibilities of the vacua, but local minimum and intrinsic
degenerate vacua are forbidden in the IR region.

\section{Supersymmetric vacua on gradient K\"ahler--Ricci soliton}\label{SUSYKRS}

In this section we consider the deformation of the
supersymmetric vacua on $U(n_c)$ symmetric gradient
K\"ahler--Ricci soliton on line bundles over
$\;{\mathrm{\lC P}}^{n_c -1}$. The organization of this
section is the following: In the first part, we review the
construction of the gradient K\"ahler--Ricci soliton on
line bundles over $\;{\mathrm{\lC P}}^{n_c -1}$ studied in
\cite{FIK}. The second part is to provide an analysis of
the supersymmetric ground states in $N=1$ theory.

\subsection{Rotationally symmetric gradient K\"ahler--Ricci soliton}

First of all, let us first construct a K\"ahler metric on $
\,{\lC}^{n_c} \backslash \{0 \}$ as follows. We define the
initial K\"ahler potential, i.e.,~at~$\tau =0$,
\begin{equation}\label{Kpotans}
K(z, \bar{z},0) =  \phi(u),
\end{equation}
where
\begin{equation}\label{newu}
u \equiv 2\ln(\delta_{i\bar{j}}z^i \bar{z}^{\bar{j}}) =
2\ln \vert z \vert^2,
\end{equation}
on $ \,{\lC}^{n_c} \backslash \{0 \}$. So our ansatz
(\ref{Kpotans}) has $U(n_c)$ symmetric. Secondly, we
simplify the case by taking the vector field $Y^i(0)$ to be
holomorphic and linear\footnote{This vector field can be
derived from the real function $P(z, \bar{z})$ in
(\ref{gradvec}) \cite{cao,FIK,cao1}.}
\begin{equation}\label{killvek}
Y^i(0) =  \mu z^i,
\end{equation}
where $\mu \in \lR$. Moreover, the K\"ahler potential
(\ref{Kpotans}) implies that  the metric and its inverse
have to~be
\begin{align}
g(0) &= g_{i\bar{j}}(0)\, dz^i \, d\bar{z}^{\bar{j}} =
\left[ 2 e^{-u/2}\phi_u \,\delta_{i\bar{j}} + 4
e^{-u}\left(\phi_{uu}-\frac{1}{2}\phi_u\right)
\,\bar{z}^{\bar{i}} \, z^j \right] dz^i \, d\bar{z}^{\bar{j}},\nonumber\\
g^{-1}(0) &= g^{\bar{j}i}(0)\, \bar{\partial}_{\bar{j}}\,
\partial_i = \frac{{\rm e}^{u/2}}{2\phi_u}
\left[ \delta^{\bar{j}i}-{\rm
e}^{-u/2}\,\frac{\phi_{uu}-({1}/{2})\phi_u}{\phi_{uu}}\,
\bar{z}^{\bar{j}} z^i \right]\, \bar{\partial}_{\bar{j}}\,
\partial_i,\label{metricans}
\end{align}
where $\phi_u \equiv \frac{d\phi}{du}$ and $\phi_{uu} \equiv
\frac{d^2\phi}{du^2}$. Since the metric (\ref{metricans}) is
positive definite in this case, we have then the inequalities
\begin{equation}
\phi_u > 0, \quad \phi_{uu} > 0.\label{inequal}
\end{equation}
 Inserting (\ref{killvek}) and (\ref{metricans}) into
 (\ref{initialgeom})
 results~in
\begin{align}
&\frac{2(n_c +1)}{\phi_u}
\left(\phi_{uu}-\frac{1}{2}\phi_u\right) + \frac{4}{\phi_u}
 \left(-\left(\phi_{uu}-\frac{1}{2}\phi_u\right)+ \frac{d}{du}
 \left(\phi_{uu}-\frac{1}{2}\phi_u\right)\right) \nonumber\\
&\quad\qquad- \frac{4}{\phi_u \phi_{uu}}
\left(\phi_{uu}-\frac{1}{2}\phi_u\right)\frac{d}{du}\left(\phi_{uu}-\frac{1}{2}\phi_u\right)
+ 4 \Lambda \phi_u - 8 \mu  \,\phi_{uu}\notag\\
&\qquad= A_0 {\rm e}^{(1-n_c)u/2},
\end{align}
where $A_0$ is an arbitrary constant. Now defining $\Phi
\equiv \phi_u$ and then writing $\Phi_u = F(\Phi)$ yields
\begin{equation}\label{FPhi}
\frac{dF}{d\Phi}+\left(\frac{n_c -1}{\Phi}-4\mu\right)F -
\left(\frac{n_c }{2}-2\Lambda \Phi\right) = \frac{A_0}{\Phi} {\rm
e}^{(1-n_c)u/2}.
\end{equation}
For the sake of simplicity, we consider the case where the
right hand side of  (\ref{FPhi}) vanishes by taking $A_0
=0$, so that the solution of (\ref{FPhi}) takes the~form
\begin{equation}\label{solFPhi}
\Phi_u = F(\Phi) = A_1 \frac{{\rm e}^{4\mu\Phi}}{\Phi^{n_c
-1}} + \frac{\Lambda}{2\mu}\Phi  +  \frac{2(\Lambda
-\mu)}{(4\mu)^{n_c +1}}\sum_{j=0}^{n_c -1} \frac{n_c!}{j!}
(4\mu)^j \,\Phi^{j+1-n_c},
\end{equation}
where $A_1$ is also an arbitrary constant. For the case at
hand, the K\"ahler--Einstein solution of (\ref{FPhi}),
namely the $\mu =0$ case, is excluded.

To construct the soliton, let us first recall the flow
vector $X^i(\tau)$ defined in (\ref{killdiff}). From
(\ref{killvek}), we~have
\begin{equation}\label{killdifvek}
X^i(\tau) =  \frac{\mu}{\sigma(\tau)} z^i,
\end{equation}
which induces the diffeomorphisms
\begin{equation}\label{diffeo3}
\hat{z} \equiv \psi_{\tau}(z) =
\sigma(\tau)^{-\mu/2\Lambda}z.
\end{equation}
Thus, the complete $\tau$-dependent K\"ahler--Ricci soliton
defined on  $ \,{\lC}^{n_c} \backslash \{0 \}$~is
\begin{equation}\label{KRsol}
 g(z,\bar{z};\tau) = \sigma(\tau)^{1-\mu/\Lambda}
 g_{i\bar{j}}\left(\sigma(\tau)^{-\mu/2\Lambda}z \right)
 dz^i\,d\bar{z}^{\bar{j}}.
\end{equation}

Now we are ready to discuss the construction of the
gradient K\"ahler--Ricci soliton on line bundles over
$\,{\mathrm{\lC P}}^{n_c -1}$. First of all, the metric
(\ref{metricans}) is nontrivial $U(n_c)$-invariant gradient
K\"ahler--Ricci metric defined on $( \,{\lC}^{n_c}
\backslash \{0 \})/{\mathbb{Z}}_{\ell}$, where
${\mathbb{Z}}_{\ell}$ acting on $\,{\lC}^{n_c} \backslash
\{0 \}$ by $z  \mapsto e^{2\pi {\mathrm{i}}/\ell}z$ with
$\ell$ is a positive integer and $\ell \ne 0$. Thus, we
replace the coordinates $z$ using the above analysis by new
coordinates which can be viewed as holomorphic polynomial
of order $\ell$,~i.e.,
\begin{equation}\label{newcoor}
\xi \equiv z^{\ell},
\end{equation}
 that parameterize $(\,{\lC}^{n_c} \backslash \{0
 \})/{\mathbb{Z}}_{\ell}$.
 Then we construct a negative line bundle over $\,{\mathrm{\lC P}}^{n_c -1}$,
 denoting by $L^{-\ell}$,
  by gluing $\,{\mathrm{\lC
P}}^{n_c -1}$ into $( \,{\lC}^{n_c} \backslash \{0
\})/{\mathbb{Z}}_{\ell}$ at\break $\xi=0$.\footnote{We do
not consider $L^{\ell}$ for $\ell
>0$ here because there is no complete gradient
K\"ahler--Ricci soliton on it.} This leads that one has to
add $\,{\mathrm{\lC P}}^{n_c -1}$ at $\xi =0$ and analyze
(\ref{solFPhi}) near $\xi =0$. In this case, it is
convenient to cast the initial metric (\ref{metricans})
into the
 following~form
\begin{equation}
 g(0) = \Phi \, g_{\rm FS} + \Phi_v \, dw\,d\bar{w},
\end{equation}
where $g_{\rm FS}$ is the standard Fubini--Study metric of
$\,{\mathrm{\lC P}}^{n_c -1}$
\begin{equation}\label{FSmetric}
g_{\rm FS} = \left( \frac{\delta_{a\bar{b}}}{1+\zeta^c
\bar{\zeta}^{\bar{c}}}-\frac{\zeta^b
\bar{\zeta}^{\bar{a}}}{(1+\zeta^c
\bar{\zeta}^{\bar{c}})^2}\right)d\zeta^a\,d\bar{\zeta}^{\bar{b}},
\end{equation}
 with $\zeta^a \equiv \xi^a /\xi^{n_c}$, $\zeta^{n_c} =1$ and $a,b,c = 1,\ldots,n_c -1$.
 Moreover, $w \equiv w(\xi,\bar{\xi})$ is a
nonholomorphic coordinate and here $v \equiv 2\ln \vert \xi
\vert^2$. So, to obtain $\,{\mathrm{\lC P}}^{n_c -1}$ at
$\xi =0$, one has to take
\begin{equation}
 \lim_{v \to -\infty} \Phi(v) = a > 0, \quad F(a)= 0,
 \quad \frac{dF}{d\Phi}(a) > 0
\end{equation}
with
\begin{equation}
a = \frac{1}{ 4 \Lambda } (n_c - \ell).
\end{equation}
For $\Lambda < 0$ case, we have $\ell > n_c$, whereas the
$\Lambda
> 0$ case demands $0< \ell < n_c$.\footnote{In \cite{FIK} for
$\Lambda < 0$ case the metric (\ref{KRsol}) becomes
asymptotically cone-like expanding soliton, while in the
case of $\Lambda > 0$  it turns out to be cone-like
shrinking soliton in the asymptotic region.}

Let us consider the behavior of the gradient
K\"ahler--Ricci soliton defined on $L^{-\ell}$ in the
asymptotic region $\vert \xi \vert \to +\infty$. We take a
case where $\Phi$ becomes larger as $v \to +\infty$ such
that the exponential term in (\ref{solFPhi}) suppresses to
zero and the dominant term is the linear term
$\frac{\Lambda}{2\mu}\Phi$. The inequalities
(\ref{inequal}) further restricts that we only have two
possible solutions as follows. The first case is when
$\Lambda <0$ and $\mu < 0$. Then (\ref{solFPhi}) may be
written down in the~form
\begin{equation}\label{solinex}
\Phi_v = \frac{\Lambda}{2\mu}\, \Phi  +
G\left(\frac{1}{\Phi}\right)\!.
\end{equation}
Rewriting (\ref{solinex}) in terms of $\Psi = 1/\Phi$
results~in
\begin{equation}\label{solexp1}
\Phi = {\rm e}^{\Lambda v/2\mu} B\left({\rm e}^{-\Lambda
v/2\mu}\right)\!,
\end{equation}
for large $v$, where $B$ is smooth and $B(0)>0$. Thus, the
K\"ahler--Ricci soliton (\ref{KRsol}) becomes
\begin{align}
 g(\xi,\bar{\xi};\tau) &=2\Bigg[\vert \xi \vert^{-2+ 2 \Lambda /\mu}\,
 B\left(\sigma(\tau)
 \vert \xi \vert^{-2 \Lambda /\mu}\right) \delta_{i\bar{j}}\nonumber\\
  &\quad+ \Bigg\{\left(\frac{\Lambda}{\mu}-1\right)\,
  \vert \xi \vert^{2 \Lambda /\mu} B\left(\sigma(\tau)
 \vert \xi \vert^{-2 \Lambda /\mu}\right) \nonumber\\
&\quad- \sigma(\tau)\, \frac{\Lambda}{\mu} \,
\dot{B}\left(\sigma(\tau)
 \vert \xi \vert^{-2 \Lambda  /\mu}\right)\Bigg\} \vert \xi \vert^{-4}
 \bar{\xi}^{\bar{i}} \, \xi^j \Bigg] d\xi^i\,d\bar{\xi}^{\bar{j}}, \label{conexsol}
\end{align}
with $\xi \in L^{-\ell}$. It is easy to see that the soliton
(\ref{conexsol}) collapses to a cone metric at $\tau = \frac{1}{
2 \Lambda } < 0$
\begin{equation}
g(\xi,\bar{\xi};1/2\Lambda) = B(0) \,
g_{\mathrm{cone}}(\xi,\bar{\xi}),
\end{equation}
where
\begin{equation}\label{conemet}
 g_{\mathrm{cone}}(\xi,\bar{\xi}) =  \vert \xi \vert^{-2+ 2 \Lambda /\mu}\,
 \left(  \delta_{i\bar{j}}
   + \left(\frac{\Lambda}{\mu}-1\right)
   \vert \xi \vert^{-2}
 \bar{\xi}^{\bar{i}} \, \xi^j  \right) d\xi^i\,d\bar{\xi}^{\bar{j}},
\end{equation}
and then expands smoothly for $\tau \ge 0$.

Now we consider the second case where $\Lambda
>0$ and $\mu >0$. In this case, as $\Phi$ becomes larger
as $v \to +\infty$, the exponential term becomes dominant
and then it is difficult to define $\Phi$ in this region.
To get rid of it, one has to set $A_1 =0$, which further
implies that the dominant term is again the linear term
$\frac{\Lambda}{2\mu}\Phi >0$. Then, we may write
(\ref{solFPhi}) into (\ref{solinex}), which leads to a
similar equation like (\ref{solexp1}).  After some
computation, we get
\begin{align}
 g(\xi,\bar{\xi};\tau) &=2\Bigg[\vert \xi \vert^{-2+ 2 \Lambda /\mu}\,
  D\left(\sigma(\tau)
 \vert \xi \vert^{-2 \Lambda /\mu}\right) \delta_{i\bar{j}}\nonumber\\
  &\quad+ \Bigg\{\left(\frac{\Lambda}{\mu}-1\right)\,
  \vert \xi \vert^{2 \Lambda /\mu} D\left(\sigma(\tau)
 \vert \xi \vert^{-2 \Lambda /\mu}\right) \nonumber\\
&\quad- \sigma(\tau)\, \frac{\Lambda}{\mu} \,
\dot{D}\left(\sigma(\tau)
 \vert \xi \vert^{-2 \Lambda  /\mu}\right)\Bigg\} \vert \xi \vert^{-4}
 \bar{\xi}^{\bar{i}} \, \xi^j \Bigg] d\xi^i\,d\bar{\xi}^{\bar{j}},  \label{conshsol}
\end{align}
for large $\vert \xi \vert$, where $D$ is a smooth function
and $D(0) >0$ with $\xi \in L^{-\ell}$. Therefore, the
metric (\ref{conshsol}) converges to a cone metric
(\ref{conemet}) at $\tau = \frac{1}{ 2 \Lambda} > 0$ and
evolves smoothly for~$\tau > \frac{1}{ 2 \Lambda}$.

\subsection{Vacuum structure of $N=1$ theory}

\indent First of all, we recall some properties of
Minkowskian ground states of the local theory discussed in
(\ref{LSC}) since the second-order analysis is similar to
the global case by setting $M_P \to +\infty$. As pointed
out before, a flat Minkowskian vacuum $\tilde{p}_0 \equiv
(\xi_0, \bar{\xi}_0)$ requires
\begin{equation}\label{singMink}
W(\xi_0)= \partial_i W(\xi_0) = 0,
\end{equation}
 together with their complex conjugate for all $i=1,\ldots,n_c$.
 Again, we obtain here a static Minkowskian vacuum which is
 defined by the critical points of~the~holomorphic superpotential.
 The nonzero component of the Hessian matrix of the scalar
potential (\ref{V1}) is given~by
\begin{equation}
 \partial_i \bar{\partial}_{\bar{j}}{\mathcal{V}}(\tilde{p}_0 ; \tau)
 = {\rm e}^{K(\tilde{p}_0 ; \tau)/M_P^2}
 g^{l\bar{k}}(\tilde{p}_0 ;\tau)\, \partial_i
\partial_l W(\xi_0) \,
\bar{\partial}_{\bar{k}}\bar{\partial}_{\bar{j}}\bar{W}(\bar{\xi}_0),
\end{equation}
where $K(\tau)$ and $g^{l\bar{k}}(\tau)$ are the K\"ahler
potential and the inverse of the metric (\ref{KRsol}),
respectively, with vanishing scalar potential. We first
assume that the scalar potential (\ref{V1}) is a Morse
potential, so the holomorphic superpotential has at least
 quadratic form with invertible Hessian matrix discussed in Section
 \ref{GSC}. Near
$\xi =0$, the analysis is the same as in Section
\ref{SUSYKEM} since we have $\;{\mathrm{\lC P}}^{n_c -1}$,
which is a K\"ahler--Einstein manifold. Therefore, our
interest is to consider in asymptotic region $\vert \xi
\vert \to +\infty$.

For the case at hand, we take the case for $\tau \ge 0$.
So, in $\Lambda < 0$ case no collapsing point exists and
the soliton expands smoothly, described by
(\ref{conexsol}), whose $U(1)$-connection and K\"ahler
potential of the theory are given~by
\begin{align}
 Q(\xi,\bar{\xi};\tau) &=  -\frac{\mathrm{i}}{M_P^2} \vert \xi
 \vert^{-2+ 2 \Lambda /\mu}\,
 B\left(\sigma(\tau) \vert \xi \vert^{-2 \Lambda /\mu}\right)
 \left[ \bar{\xi}^{\bar{i}} \, d\xi^i - \xi^i \, d\bar{\xi}^{\bar{i}}
 \right], \nonumber\\
K(\xi,\bar{\xi};\tau) &= \int {\rm e}^{\Lambda v/2\mu}
B\left(\sigma(\tau)e^{-\Lambda v/2\mu} \right) dv +
c,\label{U1conex}
\end{align}
respectively, where $v \equiv 2\ln \vert \xi \vert^2$
 and $c$ is a real constant. Since $B$ is
smooth function and, in general, positive definite as
demanded by the inequalities (\ref{inequal}), then it is
well defined and does not have other singular point,
namely, for $\tau = \tau_0 \ge 0$ such~that
\begin{equation}\label{Bsmooth}
B\left(\sigma(\tau_0) \vert \xi \vert^{-2 \Lambda /\mu}\right) =
\dot{B}\left(\sigma(\tau_0)
 \vert \xi \vert^{-2 \Lambda  /\mu}\right) =0.
\end{equation}

\removelastskip\pagebreak

\noindent This can be restated that it is impossible to
have $\tau_0 \ge 0$ at which the soliton (\ref{conexsol})
collapses to a point. On the other hand, in $\Lambda > 0$
case, the soliton (\ref{conshsol}) converges to a cone
metric at $\tau = 1/2\Lambda$ which is singular manifold
because the diffeomorphism (\ref{diffeo3}) diverges. The
function $D$ should have the same behavior as $B$, i.e., it
does not satisfy (\ref{Bsmooth}) for $\tau \ge 0$.
Moreover, since $\vert \xi \vert \to +\infty$, and then, in
order to achieve a consistent picture we have to take
$\vert \xi \vert \gg \tau$ for any finite $\tau$.
Therefore, the solitons (\ref{conexsol}) and
(\ref{conshsol}) can be expanded as
\begin{equation}\label{conedominance}
g(\xi,\bar{\xi};\tau) = g_{\mathrm{cone}}(\xi,\bar{\xi}) +
O(\xi,\bar{\xi}; \tau),
\end{equation}
where $O(\xi,\bar{\xi}; \tau)$ is higher order terms which
are very small compared to
$g_{\mathrm{cone}}(\xi,\bar{\xi})$ given in
(\ref{conemet}). Hence, in both cases the solitons are
invertible, positive definite, and correspondingly, we have
well-defined theory for all $\tau \ge 0$. Further analysis
confirms that there is no parity transformation of the
Hessian matrix of (\ref{V1}) that changes the Morse index
of Minkowskian ground states.

Now we turn our attention to discuss nondegenerate AdS
vacuum. In this case, supersymmetry further requires
\begin{equation}\label{susyconGKR}
 \partial_i W(\xi_0) +
 \frac{2}{M_P^2} \vert \xi_0 \vert^{-2+ 2 \Lambda /\mu}\,
 B\left(\sigma(\tau) \vert \xi_0 \vert^{-2 \Lambda /\mu}\right)
  \bar{\xi}_0^{\bar{i}} \, W(\xi_0) =0
\end{equation}
for $\Lambda <0$, which follows that the ground state also
depends on $\tau$, namely $\tilde{p}_0(\tau)$ with non zero
scalar potential
\begin{equation}\label{VAdS1}
{\mathcal{V}}(\tilde{p}_0; \tau) =  - \frac{3}{M_P^2} {\rm
e}^{K(\tilde{p}_0; \tau)/M_P^2}\vert W(\xi_0)\vert^2,
\end{equation}
where the K\"ahler potential $K(\tau)$ is given in
(\ref{U1conex}), while for $\Lambda >0$ the function $B$ is
replaced by $D$. However, in the case at hand, since we
have cone dominating metric, only static AdS vacuum leads
the term in (\ref{susyconGKR}) and further,
\begin{align}
 B\left(\sigma(\tau) \vert \xi \vert^{-2 \Lambda /\mu}\right)
 &\approx B(0), \nonumber\\
K(\xi,\bar{\xi};\tau) &\approx B(0) \frac{2 \mu}{\Lambda}
\, \vert \xi \vert ^{2\Lambda / \mu} + c,
    \label{U1conex1}
\end{align}
for any $ \xi $ in this asymptotic region. In addition, it
is impossible to find a vacuum $\tilde{p}_0$ defined by the
critical points of $W(z)$ as in the K\"ahler--Einstein case
in which the $U(1)$-connection in (\ref{U1conex}) evaluated
at $\tilde{p}_0$ vanishes.

For degenerate cases, both Minkowskian and AdS vacua reach
the same results, namely the parity transformation of the
Hessian matrix of (\ref{V1})\vadjust{\pagebreak} caused by
the soliton does not emerge because the static cone metric
$g_{\mathrm{cone}}\break(\xi,\bar{\xi})$ is the leading
term. Hence, we obtain similar results as in the previous
case.

Similar as in the preceding section, the next step is to
check the existence of such AdS vacua using (\ref{beta1})
describing the first-order expansion of the beta function
(\ref{beta}). For example, in the model where all
spin-$\frac{1}{2}$ fermions are massless, all AdS vacua
live in the UV region because all the eigenvalues of
(\ref{beta1}) are positive. Moreover, such a model admits
only unstable walls because only negative eigenvalues of
the Hessian matrix of (\ref{V1})\break survive.

Finally, we conclude all results by the following theorem.

\newtheorem{Dinmorse}{Theorem}[section]

\begin{Dinmorse}\label{Dinmorselemma}
Consider $N=1$ chiral supersymmetry on $U(n_c)$ symmetric
K\"ahler--Ricci soliton {\rm (\ref{KRsol})} in asymptotic
region satisfying {\rm (\ref{conedominance})}. Suppose
there exists an $m$-dimensional vacuum manifold $S$ of the
theory with the index $\lambda$ in the IR or UV regions.
Then, in the normal direction of $S$, we can introduce real
local coordinates $X_p(\tau) \approx X_p$ with $p=
1,\ldots, 2n_c -m$ for all $\Lambda$ and finite $\tau$
such~that
\begin{equation}\label{VMink}
{\mathcal{V}}(\tau) = {\mathcal{V}}(S) -
X_1^2-\cdots-X_{\lambda}^2
 +X_{\lambda +1}^2+\cdots+X_{2n_c -m}^2.
\end{equation}
Thus, no parity transformation of the Hessian matrix of
(\ref{V11}) emerges caused by the K\"ahler--Ricci soliton
in the region.
\end{Dinmorse}

\section{Conclusions}\label{concl}

So far we have discussed the properties of four-dimensional
chiral $N=1$ supersymmetry on K\"ahler--Ricci flow
satisfying (\ref{KRF}). Both in global and local theories,
all couplings deform with respect to the parameter $\tau$
which is related to the dynamics of the K\"ahler metric.
Then, flat BPS domain walls of $N=1$ supergravity have been
constructed where the warped factor (\ref{warp}) and the
supersymmetric flows are described by the BPS equations
(\ref{gfe}) and the beta function (\ref{beta}) is
controlled by the holomorphic superpotential $W(z)$ and the
geometric flow (\ref{KRF}).

We have considered the case where the initial manifold is
K\"ahler--Einstein with $\Lambda > 0$. In particular, in
the interval $\tau \ge 0$, the flow collapses to a point at
$\tau = 1/2\Lambda$ and both $N=1$ theories diverge. Such
singularity disjoints two different theories, namely in $0
\le \tau < 1/2\Lambda$ we have $N=1$ theories on a
K\"ahler--Einstein manifold with $\Lambda
>0$, whereas for $\tau > 1/2\Lambda$ the case\vadjust{\pagebreak} turns out to be
$N=1$ theories on another K\"ahler--Einstein manifold with
$\Lambda < 0$ and \hbox{opposite} metric signature. In the
latter model we have wrong-sign kinetic term which might be
considered as ghosts.

In global theory, such geometrical deformation affects the
vacua, which can be directly seen from the parity
transformation of the Hessian matrix (\ref{HessV}) that
changes the Morse index of the vacua if they are
nondegenerate. The same conclusion is achieved for
nondegenerate Minkowskian vacua in the local theory. These
are written in Theorem~\ref{Modmorselemma}.

In AdS vacua, it is not trivial to observe directly
 such phenomenon
 since both the geometric flow (\ref{KRF}) and the coupling quantities
 such as the holomorphic superpotential $W(z)$  and the fermionic mass in
 (\ref{fermass1}) generally determine the properties
  of the vacua in the second-order analysis. Therefore, we had
  to impose the conditions (\ref{mirrorcon}) in order for
 the parity transformation to be revealed in both the nondegenerate
  and degenerate cases
 described by Theorem \ref{LemmaVmorse1} and Theorem \ref{LemmaVmorse2},
 respectively. Particularly in the degenerate case we had another possibility
 that the K\"ahler--Ricci flow could also change an $n$-dimensional
 vacuum manifold to another $m$-dimensional vacuum manifold with $n \ne m$
  as mentioned in Theorem \ref{LemmaVmorse3}. Finally, we have employed the
  RG flow analysis to verify the existence of the vacua since
  they correspond to the three dimensional~CFT.

Then, we gave an explicit model, namely, the
$\;{\mathrm{\lC P}}^{n_c}$
  model with linear superpotential $W(z) = a_0 + a_i \, z^i$.
 In the model, we obtain a static vacuum $z^i =0$ with $a_i =0$.
 For $a_0 \ne 0$,  the vacuum is an isolated AdS spacetime of the index
 $2n_c$ describing unstable walls for $0\le \tau < 1/2(n_c +
1)$. In the interval $ \tau > 1/2(n_c + 1)$, the index then
changes to $0$ and it corresponds to stable walls. All of
them live in the UV region ensured by (\ref{beta2}). On the
other side, taking $a_0 =0$ for any $z^i$ the vacua become
degenerate Minkowskian spacetime. Hence, the vacuum
manifold is $\;\lC P^{n_c}$ for $0 \le \tau < 1/2 (n_c +1)$
and then turns into $\;\widehat{\lC {\rm P}^{n_c}}$ for~$
\tau > 1/2(n_c +1)$.

\enlargethispage{6pt}

For dynamical AdS vacua, we easily take $n_c =1$. Before
singularity, namely $\tau < 1/4$, there are local minimum,
local maximum, saddle, and degenerate vacua living in the
UV region given by (\ref{locmin}) to (\ref{degtau}),
whereas in the IR region we have only local minimum,
saddle, and nonintrinsic degenerate vacua. This evidence
shows that the K\"ahler--Ricci flow indeed plays a role in
changing the dimension and the index of the vacua
consistent with the first point in Theorem
\ref{LemmaVmorse3}. Similar phenomenon happens after
singularity, namely $\tau > 1/4$. The UV region allows all
possibilities of vacua, but in the IR region there are only
local maximum, saddle, and again, nonintrinsic degenerate
vacua. These are the facts of the second point in\break
Theorem~\ref{LemmaVmorse3}.

The second example is the $U(n_c)$ invariant soliton.
Around $\xi =0$ the soliton becomes $\;{\mathrm{\lC
P}}^{n_c -1}$, which is a K\"ahler--Einstein manifold
mentioned above. Therefore, our interest is in asymptotic
region $\vert \xi \vert \to +\infty$. Again, for $\Lambda
>0$, the singularity occurs at $\tau = 1/2\Lambda$ and the
soliton converges to a K\"ahler cone. Additionally, the
soliton has positive definite metric for $\tau \ge 0$
demanded by the inequalities~(\ref{inequal}).

Moreover, in order to get a consistent picture we have to
choose $\vert \xi \vert >> \tau$ for any finite $\tau$.
Then, in the region the soliton is dominated by the cone
metric (\ref{conemet}) and has positive definite for all
$\tau  \ge 0$. Therefore, all vacua are static and
correspondingly, do not posses parity transformation of the
Hessian matrix of the scalar potential (\ref{V1}) caused by
the geometric flow (\ref{conexsol}) and (\ref{conshsol}).
The same step as in the previous case, in AdS vacua the
analysis using RG flow must be performed for proving the
existence of such vacua. As an example, in a model
consisting of all massless spin-$\frac{1}{2}$ fermions, all
AdS vacua exist in the UV region.

Finally, we want to mention that the analysis here can
 be generalized to a case of curved BPS domain walls.
This has been addressed for two dimensional model
in~\cite{GYZ}.

\section*{Acknowledgment}

We acknowledge H.-D. Cao and T. Mohaupt for email
correspondence. It is also a pleasure to thank K. Yamamoto
and the people at the theoretical astrophysics group and
the elementary particle physics group of Hiroshima
university for nice discussions and warmest hospitality
during the authors stay in Theoretical Astrophysics Group,
Hiroshima University. We also thank M. Nitta for notifying
us the references \cite{HI, MN}. We would like to thank M.
Satriawan for reading the previous version of the paper and
correcting some grammar. This work is funded by ITB Alumni
Association (HR IA-ITB) research project 2008 No.
1241a/K01.7/PL/2008, Riset Fundamental DP2M-DIKTI
No.013/SP2H/PP/DP2M/III/2007, and
324/SP2H/PP/DP2M/III/2008.

\appendix

\section*{A\quad Convention and notation}

The aim of this appendix is to assemble our conventions in
this paper. The spacetime metric is taken to have the
signature $(+,-,-,-)$ while the Riemann tensor is defined
to be $R^{\mu}_{\;
\nu\rho\lambda}=\partial_{\rho}\Gamma^{\mu}_{\;\nu\lambda
}-\partial_{\lambda}\Gamma^{\mu}_{\;\nu\rho} +
\Gamma^{\sigma}_{\;\nu\lambda}\Gamma^{\mu}_{\;\sigma\rho} -
\Gamma^{\sigma}_{\;\nu\rho}\Gamma^{\mu}_{\;\sigma\lambda}$.
The Christoffel symbol is given by $\Gamma^{\mu}_{\;
\nu\rho}=\frac{1}{2}g^{\mu\sigma}(\partial_{\nu}g_{\rho\sigma}
+
\partial_{\rho}g_{\nu\sigma}-\partial_{\sigma}g_{\nu\rho})$
where $g_{\mu\nu}$ is the spacetime metric.

We supply the following indices:
\begin{align*}
\mu,\nu = 0,\ldots,3, &\quad \hbox{label curved four-dimensional spacetime indices}; \\
i,j,k=1,\ldots,n_c, &\quad  \hbox{label the } N=1 \hbox{ chiral multiplet}; \\
\bar{i},\bar{j},\bar{k} =1,\ldots,n_c, &\quad \hbox{label
conjugate indices of }
i,j,k;\\
a, b = 1,\ldots,n_c -1, &\quad \hbox{label Fubini--Study
coordinates
of } \,{\mathrm{\lC P}}^{n_c -1};\\
p,q = 1,\ldots, 2n_c, &\quad \hbox{label real coordinates};\\
\underline{p}, \underline{q} = 1,\ldots, 2n_c-m, &\quad
\hbox{label real
coordinates in normal direction} \\
&\quad \hbox{of } m\hbox{-}\hbox{dimensional submanifold}.
\end{align*}

Some quantities on a K\"ahler manifold are given as
follows. $g_{i\bar{j}}$ denotes the metric of the K\"ahler
manifold whose Levi--Civita connection is defined as
$\Gamma^l_{\;ij} = g^{l\bar{k}}\partial_i g_{j\bar{k}}$ and
its conjugate $\Gamma^{\bar{l}}_{\;\bar{i}\bar{j}} =
g^{\bar{l}k}\partial_{\bar{i}} g_{\bar{j}k}$. Then the
curvature of the K\"ahler manifold is
defined~as\def\theequation{A.\arabic{equation}}\setcounter{equation}{0}
\begin{equation}
R^l_{\;i\bar{j}k} \equiv \bar{\partial}_{\bar{j}}
\Gamma^l_{\;ik},
\end{equation}
while the Ricci tensor has the form
\begin{equation}
R_{k\bar{j}} =  R_{\bar{j}k} \equiv R^i_{\;i\bar{j}k} =
\bar{\partial}_{\bar{j}} \Gamma^i_{\;ik} = \partial_k
\bar{\partial}_{\bar{j}}\ln\left(
{\mathrm{det}}(g_{i\bar{j}})\right)\!.
\end{equation}
Finally, the Ricci scalar can be written down as
\begin{equation}
R \equiv g^{i\bar{j}}R_{i\bar{j}}.
\end{equation}

\section*{B\quad Analysis of the Hessian matrix of the scalar potential}\label{AHMSP}

This section is set to give our convention of the Hessian
matrix of the scalar potential and some proofs of the
results particularly discussed in Section~\ref{SUSYKEM}
since the other results can be extracted from these proofs.
The analysis here is standard in Morse(-Bott) theory and we
omit the Einstein summation convention. Interested reader
can further read references~\cite{JM,YM,BH,BH1}.

First of all, we define the Hessian matrix of the scalar
potential (\ref{V1}) in local
theory~as\def\theequation{B.\arabic{equation}}\setcounter{equation}{0}
\begin{equation}\label{HessV3}
H_{\mathcal{V}}  \equiv  \left(
\begin{matrix}
\partial_i \bar{\partial}_{\bar{j}}{\mathcal{V}}
   & & \partial_i \partial_j {\mathcal{V}} \\
  & & \\
  \partial_j \partial_i {\mathcal{V}}
  & & \partial_j \bar{\partial}_{\bar{i}}{\mathcal{V}} \\
\end{matrix}
\right) (p_0),
\end{equation}
where $p_0 = (z_0, \bar{z}_0)$ is a critical point (vacuum)
and $i,j =1,\ldots,2n_c$, around which the scalar potential
can be expanded~as
\begin{equation}\label{Vgen}
{\mathcal{V}}(z, \bar{z}; \tau) = {\mathcal{V}}(p_0 ; \tau)
+ \sum_{p,q=1}^{2n_c} \frac{\partial^2
{\mathcal{V}}(p_0)}{\partial x^p \partial x^q} \, \delta
x^p \, \delta x^q,
\end{equation}
with $p,q = 1,\ldots,2n_c$, and we have defined real
coordinates such that $z^i \equiv x^i + {\mathrm{i}}
x^{i+n_c}$ and $\delta x^p \equiv x^p - x_0^p$.

Let us first focus on nondegenerate case and the initial
geometry is K\"ahler--Einstein with $\Lambda >0$. For the
case at hand, the matrix (\ref{HessV3}) is invertible and
evolves~as
\begin{equation}\label{HessV4}
H_{\mathcal{V}}  = \varepsilon(\sigma) \left(
\begin{matrix}
 {\mathcal{V}}_{i\bar{j}}(p_0)
   & & {\mathcal{V}}_{ij}(p_0) \\
  & & \\
  {\mathcal{V}}_{ji}(p_0)
  & & {\mathcal{V}}_{j\bar{i}}(p_0) \\
\end{matrix}
\right)\!,
\end{equation}
where $\varepsilon(\sigma)$ is given in (\ref{varep}) and
the quantities ${\mathcal{V}}_{i\bar{j}}(p_0)$,
${\mathcal{V}}_{ij}(p_0)$ are defined in (\ref{Vij}). In
order to obtain the parity transformation of
(\ref{HessV3}), the real and imaginary parts of
${\mathcal{V}}_{i\bar{j}}(p_0)$ and
${\mathcal{V}}_{ij}(p_0)$ remain positive with respect to
$\tau \ge 0$ and $\tau \ne 1/2\Lambda$, which means that
the inequalities (\ref{mirrorcon}) hold. Therefore, the
expansion (\ref{Vgen}) becomes
\begin{equation}\label{VgenKE}
{\mathcal{V}}(z, \bar{z}; \tau) = {\mathcal{V}}(p_0(\tau);
\tau) + \varepsilon(\sigma) \sum_{p,q=1}^{2n_c}
{\mathcal{V}}_{pq}(p_0(\tau); \tau) \, \delta x^p \, \delta
x^q,
\end{equation}
with $x^p_0\equiv x^p_0(\tau)$. Since $\tau \ne
1/2\Lambda$, we can define a new coordinate $X_1(\tau)$~as
\begin{equation}\label{newcoorX}
X_1(\tau) \equiv \vert {\mathcal{V}}_{11}(p_0(\tau);
\tau)\vert^{1/2} \left( \delta x^1(\tau) +
\sum_{p=2}^{2n_c} \delta x^p(\tau)
\frac{{\mathcal{V}}_{p1}(p_0(\tau)}{{\mathcal{V}}_{11}(p_0(\tau)}
\right)\!,
\end{equation}
and then, (\ref{VgenKE}) can be expressed as
\begin{equation}\label{VgenKE1}
{\mathcal{V}}(z, \bar{z}; \tau) = {\mathcal{V}}(p_0(\tau);
\tau) + \varepsilon(\sigma) \left (\pm \, X^2_1(\tau) +
\sum_{p,q=2}^{2n_c}{\mathcal{V}}'_{pq}(p_0(\tau); \tau) \,
\delta x^p \, \delta x^q \right)\!.
\end{equation}
Therefore, we inductively carry on the computation until $r <
2n_c$ by defining
\begin{equation}\label{newcoorXr}
X_r(\tau) \equiv \vert
\widetilde{\mathcal{V}}_{rr}(p_0(\tau); \tau)\vert^{1/2}
\left( \delta x^r(\tau) + \sum_{p = r+1}^{2n_c} \delta
x^p(\tau) \frac{\widetilde{\mathcal{V}}_{pr}(p_0(\tau);
\tau)}{\widetilde{\mathcal{V}}_{rr}(p_0(\tau); \tau)}
\right)\!,
\end{equation}
so that (\ref{VgenKE}) has the form
\begin{equation}\label{VgenKEr}
{\mathcal{V}}(z, \bar{z};
\tau)\,{=}\,{\mathcal{V}}(p_0(\tau); \tau)\,{+}\,
\varepsilon(\sigma)\!\left(\!{\pm}\sum_{p=1}^{r}
X^2_p(\tau) \,{+}\,\sum_{p,q=r+1}^{2n_c}
\widetilde{\mathcal{V}}'_{pq}(p_0(\tau); \tau) \, \delta
x^p \, \delta x^q\!\right)\!,
\end{equation}
where
\begin{equation}
X_p(\tau) =\begin{cases}
Y_p(\tau) &   {\text{if}} \;\;\;  0 \le \tau < 1/2\Lambda, \\
\widehat{Y}_p(\tau)  &  {\text{if}} \;\qquad \; \, \tau > 1/2\Lambda.\\
\end{cases}
\end{equation}
This completes the induction and the proof of Theorem
\ref{LemmaVmorse1}. If we take the limit $M_P \to +\infty$
or $W(z_0)=0$, then $X_p(\tau)= \vert \sigma(\tau)
\vert^{-1/2} X_p(0)$ as discussed in
Theorem~\ref{LemmaVmorse1}.

Now we turn to consider the degenerate case starting with
the expansion (\ref{VgenKE}). Next, the infinitesimal dual
basis $dx^p$ are split~into\vspace*{-2pt}
\begin{equation}\label{basesplit}
 dx^p= \sum_{\underline{p}=1}^{2n_c-m}\Big(T_{\;\underline{p}}^p \; dx^{\underline{p}} + N_{\;\underline{p}}^p \;
dx^{\underline{p}}\Big).\vspace*{-2pt}
\end{equation}
where $T_{\;\underline{p}}^p \; dx^{\underline{p}}$ and
$N_{\;\underline{p}}^p \; dx^{\underline{p}}$ are both dual
tangent and normal vectors of an $m$-dimensional
submanifold $S$, respectively. Furthermore, the Hessian
matrix in the direction of the tangent vector vanishes,
namely\vspace*{-2pt}
\begin{equation}
 \sum_{p=1}^{2n_c}  T_{\;\underline{p}}^p \;
 {\mathcal{V}}_{pq}(p_0(\tau); \tau) \,
 = 0\vspace*{-2pt}
\end{equation}
for each $q$ and $\underline{p}$. Restricting to the normal
subspace of $S$, (\ref{VgenKE}) can be
written~as\vspace*{-2pt}
\begin{equation}\label{VgenKEdeg}
{\mathcal{V}}(z, \bar{z}; \tau) = {\mathcal{V}}(p_0(\tau); \tau) +
\varepsilon(\sigma) \sum_{\underline{p}, \underline{q}=1}^{2n_c-m}
 {\mathcal{V}}_{\underline{p}\underline{q}}(p_0(\tau); \tau) \,
\delta x^{\underline{p}} \, \delta
x^{\underline{q}},\vspace*{-2pt}
\end{equation}
where ${\mathcal{V}}_{\underline{p}\underline{q}}(p_0(\tau); \tau)
 \equiv \sum_{p,q=1}^{2n_c}N_{\;\underline{p}}^p N_{\;\underline{q}}^q
{\mathcal{V}}_{pq}(p_0(\tau); \tau)$. We can then do a
similar computation as in the nondegenerate case to cast
(\ref{VgenKEdeg}) in a bilinear form. Thus, we have proved
Theorem~\ref{LemmaVmorse2}.\vspace*{-3pt}

\section*{C\quad More on K\"ahler--Ricci flow}\label{MKRF}

In this appendix, we provide dynamical equations of
geometric quantities such as the inverse metric, Riemann
curvature, Ricci tensor, and finally, Ricci scalar. Using
the identity $g^{\bar{k}i} g_{j\bar{k}} = \delta^i_j$, we
find that the evolution of the inverse metric is
governed~by\def\theequation{C.\arabic{equation}}\setcounter{equation}{0}
\begin{equation}
\frac{\partial g^{i\bar{j}}(\tau)}{\partial \tau} = 2
R^{i\bar{j}}(\tau),
\end{equation}
where $R^{i\bar{j}} \equiv g^{i\bar{l}}
g^{k\bar{j}}R_{k\bar{l}}$, which results~in
\begin{equation}
\frac{\partial \Gamma^i_{\;jk}(\tau)}{\partial \tau} = -2
g^{i\bar{l}}\nabla_j R_{k\bar{l}}(\tau).
\end{equation}
Then, we obtain the dynamical equation of the Riemann curvature
\begin{equation}
\frac{\partial R^l_{\;i\bar{j}k}(\tau)}{\partial \tau} = -2
g^{l\bar{i}}\bar{\nabla}_{\bar{j}}\nabla_i
R_{k\bar{i}}(\tau),
\end{equation}
which consequently gives the evolution equation of the
Ricci tensor
\begin{equation}
\frac{\partial R_{\bar{j}k}(\tau)}{\partial \tau} = -2 \left(
 \Delta  R_{\bar{j}k}(\tau) + g^{l\bar{i}} R^{\bar{k}}_{\;\bar{j}l\bar{i}}
 R_{\bar{k}k}- g^{l\bar{i}}R^{i}_{\;l\bar{j}k}
 R_{i\bar{i}} \right)\!,
\end{equation}
with $\Delta \equiv g^{i\bar{j}} \nabla_i
\bar{\nabla}_{\bar{j}}$. Lastly, the dynamics of the Ricci
scalar is controlled~by
\begin{equation}
\frac{\partial R(\tau)}{\partial \tau} = -2 \left(
 \Delta  R(\tau) +  R^{i\bar{j}}
 R_{i\bar{j}} \right)\!.
\end{equation}

\end{document}